# Expediting hydrogen evolution through topological surface states on $Bi_2Te_3$

**Qing Qu[1,3†], Bin Liu[2†], Jing Liang[2,3], Ding Pan[2,4]* and Iam Keong Sou[1,2,3]***

**[1]Nano Science and Technology Program, The Hong Kong University of Science and Technology, Hong Kong, China.**

**[2]Department of Physics, The Hong Kong University of Science and Technology, Clear Water Bay, Hong Kong , China.**

**[3]William Mong Institute of Nano Science and Technology, The Hong Kong University of Science and Technology, Hong Kong, China**

**[4]Department of Chemistry, The Hong Kong University of Science and Technology, Hong Kong, China.**

**† These authors contributed equally to this work.**

***Corresponding authors**

# Abstract

Recently, the development of efficient and non-noble metal electrocatalysts with excellent durability for the hydrogen evolution reaction (HER) has attracted increasing attention. The exotic and robust metallic surface states of topological insulators (TIs) are theoretically predicted to enhance surface catalytic activity of overlaying catalysts, but no experimental evidence for TIs directly used as electrocatalysts has ever been reported. In this work, we fabricated the TI thin films of $Bi_2Te_3$ with different thicknesses using the molecular beam epitaxy method, and found that these thin films exhibit high electrocatalytic activity in HER. The 48 nm $Bi_2Te_3$ thin film has the best performance, which is attributed to its largest active area arising from the spiral growth mode of triangular domains as revealed by atomic force microscopy imaging. Importantly, our theoretical calculations reveal that while pure $Bi_2Te_3$ is not a good electrocatalyst, the $Bi_2Te_3$ thin films with partially oxidized surfaces or Te vacancies have high HER activity. The existence of the corresponding surface oxides on the $Bi_2Te_3$ thin films is supported by our X-ray photoelectron spectroscopy data. Particularly, we demonstrate that the topological surface states play a key role in enhancing the HER performance. Our study offers a new direction to design cost-effective electrocatalysts.

The combustion of hydrogen produces only water with zero carbon emission, so hydrogen as a clean energy carrier is of great importance to sustainable development[1]. The large-scale electrochemical production of hydrogen ($2H^+ + 2e^- \rightarrow H_2$) has been long hindered by the lack of a cost-effective replacement for Pt and the fact that non-precious catalytic materials are often limited by either a low catalytic efficiency or a short lifetime[2,3]. Topological insulators (TIs) are highly promising materials for various applications because of their fascinating band structures[4-7]; they are insulating in the bulk, but contain conducting surface states[8,9]. The time-reversal symmetry and particle number conservation in TIs protect the surface states from backscattering by non-magnetic impurities[10]. As the contamination of the catalyst surface may reduce catalytic activity, topological materials may help to overcome this problem because of their robust surface states[11]. Several studies indicated that topological surface states (TSSs) are of potential use in catalysis[12-18]. For example, first-principles density functional theory (DFT) calculations suggest that TIs may work as substrates to enhance the activity of overlaying catalysts[12-14]. Experimentally, He et al. found that the TSSs from $Bi_2Te_3$ can tunnel through a thin Pd layer and significantly enhance the surface reactivity of Pd[15]. The Weyl semimetal NbP[16] and bismuth chalcogenides based TIs were reported as efficient photocatalysts for water splitting[17]. The carbon tailored semimetal MoP was also shown to exhibit highly efficient electrocatalytic hydrogen evolution reaction (HER) performances[18]. However, to the best of our knowledge, the research on TI thin film materials directly used as electrocatalysts in HER with high performance has not been reported so far.

In this work, we fabricated $Bi_2Te_3$ thin films on ZnSe/GaAs (111)B substrate by the molecular beam epitaxy (MBE) technique. These thin films exhibit low overpotential, low Tafel slope, low charge-transfer resistance and high long-term stability in HER. It was found that $Bi_2Te_3$ thin film with a specific thickness gives the optimistic performance attributed to a feature of the spiral growth mode of the triangular domain structure. Furthermore, our DFT calculations reveals that while pure $Bi_2Te_3$ is not a good electrocatalyst, the $Bi_2Te_3$ thin films with partially oxidized surfaces or Te vacancies have a high HER activity (Fig. 1). Most importantly, we demonstrate that the TSSs play a key role in enhancing the HER performance in both cases. As research in water splitting is a highly important field in green energy generation, together with the fact that the exploration of the fascinating effects of the TSSs of topological materials is one of the major focuses in condensed matter physics research today, the findings of our study offer a new direction to optimize the performance of non–noble metal electrocatalysts based on topological quantum materials.

# Results

**Structural characterizations of MBE-grown $Bi_2Te_3$ thin films.** Fig. 2 displays the high resolution X-ray diffraction (HRXRD) profile of a typical $Bi_2Te_3$ sample. As can be seen, all the $Bi_2Te_3$ layer peaks can be indexed as (0 0 *l*) direction and their measured 2θ values give a *c*-lattice parameter of ~ 30.420 Å, which closely matches the standard value of $Bi_2Te_3$ (*c*=30.483 Å). The diffraction peaks of ZnSe (111) and ZnSe (222) are overlapped with the strong peaks of GaAs (111) and GaAs (222) in this broad-scan profile due to the very small lattice mismatch (0.27%) between ZnSe and GaAs[19-22].

The cross-sectional transmission electron microscopy (TEM) images of the $Bi_2Te_3$ samples studied in this work are shown in Supplementary Fig.1. It can be seen that each of the five samples consists of a 2.0-3.7 nm ZnSe buffer layer, and the thicknesses of $Bi_2Te_3$ thin films are determined to be 5 nm, 17 nm, 34 nm, 48 nm and 61 nm. It should be noted that one should only take these values as the approximate nominal thicknesses since these samples have domain structures at their surfaces, which will be addressed later. The fast Fourier transform (FFT) patterns shown in the insets of Supplementary Fig.1 reveal that two sets of hexagonal lattice appear in these thin films, indicating the existence of twin crystals[23], attributed to the fact that (001)-oriented $Bi_2Te_3$ has a 3-fold symmetry, while the GaAs (111) substrate has a 6-fold symmetry.

Atomic force microscopy (AFM) images of the surfaces of the five $Bi_2Te_3$ samples with different magnification are shown in Fig. 3. Fig. 3a-c are the AFM images of the 5 nm $Bi_2Te_3$ thin film with increasing magnification. As shown in the inset of Fig. 3c, several triangular domains were being developed. These domains correspond to the twin crystals with two specific orientations as revealed by the FFT patterns in the insets of Supplementary Fig.1. When the nominal thickness of $Bi_2Te_3$ reaches 17 nm, as shown in Fig. 3d-f, more domains with a few steps were developed through the spiral growth mode[24,25]. As shown in Fig. 3g-i, at thickness of the 34 nm, several neighboring domains seemed to coalesce to form a larger domain with wide terraces on which new smaller domains were being developed. Fig. 3j-l are the AFM images of the thickness of the 48 nm $Bi_2Te_3$ thin film, which show that the spiral growth of most of the smaller triangular domains had been almost fully developed so that this sample has the maximum number of layers in each domain among these samples. Fig. 3m-o are the AFM images of the 61 nm $Bi_2Te_3$ thin film, in which the triangular domains seemed to be getting merged with each other, as an evidence that the further growth of $Bi_2Te_3$ occurred at the places between the domains that have been fully developed. As shown in the profile analysis of Fig. 3c, 3f, 3i and 3l, for the layers across a spiral growth path, each layer has a height approximately equal to 1 nm, in good agreement with the thickness of a single quintuple layer (QL) $Bi_2Te_3$.

**Electrocatalytic HER performance of $Bi_2Te_3$ thin films.** We have examined the electrochemical catalytic HER behaviors of the as-grown $Bi_2Te_3$ films, as well as the ZnSe buffer and an n+GaAs substrate, the results are presented in Fig. 4. Fig. 4a displays the linear sweep voltammogram (LSV) polarization curves of these active materials, as can be seen, the overpotential ($\eta$) of the $Bi_2Te_3$ thin films at a cathodic current density ($j$) of 10 mA/cm$^2$ ranged within 219-332 mV. It is well known that the molybdenum dichalcogenide ($MX_2$) are among the most promising non-noble metal catalytic materials for HER. The comparison the $\eta$ of $Bi_2Te_3$ films in this work as well as those of exfoliated $Bi_2Te_3$, $Bi_2Se_3$, $Sb_2Te_3$, $Bi_{0.5}Sb_{1.5}Te_3$ alloy and some $MX_2$ based nanosheet catalysts[26-32] are displayed in Table 1. Our 48 nm $Bi_2Te_3$ thin film exhibits the lowest $\eta$ for HER among the reported TI materials including $Bi_2Se_3$ nanosheets(~500 mV), $Sb_2Te_3$ nanosheets(~510 mV), $Bi_{0.5}Sb_{1.5}Te_3$ nanosheets(~520 mV), exfoliated $Bi_2Te_3$ nanosheets (680 mV), and it is also among the best values shown in Table 1.

The kinetics of active catalytic materials during the HER process can be revealed by Tafel slope[33]. For HER in acidic solution, two different mechanisms, Volmer-Tafel and Volmer-Heyrovsky, have been recognized to be responsible for transforming $H^+$ to $H_2$, which include three principal steps:

$$\mathrm{H^+ + e^- + * \rightarrow H^*}\text{(Volmer reaction: electrochemical hydrogen adsorption)}, \quad (1)$$

$$\mathrm{H^* + H^+ + e^- \rightarrow H_2 + *}\text{(Heyrovsky reaction: electrochemical desorption)}; \quad (2)$$

$$\mathrm{2H^* \rightarrow H_2 + 2\,*}\text{(Tafel reaction: chemical desorption)}, \quad (3)$$

where * denotes a site on the electrode surface[34]. The corresponding Tafel plots of the active materials shown in Fig. 4a are displayed in Fig. 4b. The linear portions of these plots are fitted to the Tafel equation ($\eta = b\log j + a$, where $j$ is the current density and $b$ is the Tafel slope, the fitting equations are displayed in Supplementary Fig. 2), yielding Tafel slopes of 32.11 and a range from 47.87 to 54.10 mV/dec for Pt foil and $Bi_2Te_3$ films, respectively. For 48 nm $Bi_2Te_3$, its Tafel slope of 47.87 mV/dec is the smallest among the $Bi_2Te_3$ films, which suggests that its HER occurs through the Volmer-Heyrovsky mechanism in which a fast discharge of protons is followed by the rate-determining step of Heyrovsky reaction[35,36]. $Bi_2Te_3$ thin films with thicknesses of 5nm, 17 nm, 34 nm and 61 nm show higher Tafel slopes (54.10 mV/dec, 52.98 mV/dec, 52.94 mV/dec and 50.52 mV/dec), thus they may follow either the Volmer-Heyrovsky or the Volmer Tafel mechanism, and either Heyrovsky reaction or Tafel reaction can be the rate-determining step[35,37]. The Tafel slope of our 48 nm $Bi_2Te_3$ is among the best values listed in Table 1. We attribute the high performance in HER of the 48 nm thin film to its almost fully developed triangular domains that offer maximum active areas. The exchange current densities ($j_0$) that reflect the inherent HER activity were determined by the extrapolated x-intercepts ($\eta$=0) of the fitted dash lines, as shown in Supplementary Fig. 2. The $j_0$ of the 48 nm $Bi_2Te_3$ thin film is the highest (1.74 μA $cm^{-2}$) among all $Bi_2Te_3$ thin films (while the 5nm, 17 nm, 34 nm and 61 nm $Bi_2Te_3$ thin films have $j_0$ of 0.038 μA $cm^{-2}$, 0.173 μA $cm^{-2}$, 0.404 μA $cm^{-2}$ and 0.456 μA $cm^{-2}$, respectively), which is within the same order as those of nanosheet catalysts shown in Supplementary Table 1. The corresponding log $j_0$ (A/$cm^2$) values[34] are also given in this table. Fig. 4c shows the polarization curves of the 48 nm $Bi_2Te_3$ thin film recorded before and after 5000 cycles of cyclic voltammetry (CV) treatments. As can be seen in this figure, there is an observed potential increase of only 27 mV at $j_0$=10mA $cm^{-2}$ after the CV treatments, which indicates that the electrochemical HER process does not cause substantial changes in the electrocatalytic performance of the 48 nm thin film. This slight degradation may arise from that some active components may be exfoliated during the CV treatments due to the disturbance from the evolution of $H_2$ bubbles generated on the $Bi_2Te_3$ thin-film cathode[30]. In addition, the results of chronoamperometry (CP) test performed on the 48 nm $Bi_2Te_3$ thin film and Pt foil, as shown in Fig. 4d, further manifest that the 48 nm $Bi_2Te_3$ film exhibits an outstanding long-term operational stability, comparable to that of the commercial Pt foil. In addition, the $j$ versus reaction time curve of our thin film is more stable without the sharp spikes shown in the corresponding curve of commercial Pt foil, attributed to the much smaller size and denser of the bubbles generated during the HER process in the former case. Fig. 4e and f show the results of electrochemical impedance spectroscopy (EIS) measurements performed on the $Bi_2Te_3$ thin films, ZnSe buffer and n+GaAs substrate. By applying a simplified Randles circuit model as shown in Supplementary Fig. 3, we can derive the electrolyte resistance ($R_s$) and charge transfer resistance ($R_{ct}$) of them. The results are shown in Supplementary Table 2 in which one can see that the 48nm $Bi_2Te_3$ thin film exhibits the lowest $R_{ct}$ of 3.744 Ω $cm^2$ among the five $Bi_2Te_3$ thin films, which is consistent with the observation that its $\eta$ and Tafel slope are the lowest and $j_0$ is the highest. We believe the highly efficient performance of the 48 nm $Bi_2Te_3$ thin film is likely contributed by the fact that it has the largest active area among the five films (as shown in Fig. 3), which leads to a more efficient interaction between the catalytic surfaces and electrolytes.

**First-principles calculations.** It has been shown that the HER activity for a wide variety of catalysts can be estimated by the free energy of hydrogen adsorption ($\Delta G_H$) (see Methods)[34,38]. A catalyst will have an optimal HER activity if its $\Delta G_H$ is close to 0. For example, with using the Perdew-Burke-Ernzerhof (PBE) exchange-correlation functional, $\Delta G_H$ of Pt (111) is -0.18 eV when the surface coverage of hydrogen is 1/4 monolayer[39,40]

We first considered the hydrogen adsorption on the clean $Bi_2Te_3$ (001) basal surface in conventional cell. We tried various adsorption sites, e.g., on top of the Bi or Te atoms, above the Bi-Te bond, and at the center of the Bi-Te hexagonal ring, as shown in Supplementary Fig. 4, and found that the $\Delta G_H$ for all these sites are too high. The smallest $\Delta G_H$ is 0.53 eV. Here, the positive energy indicates a repulsive force between the hydrogen atom and the $Bi_2Te_3$ slab, so it is difficult to bond the hydrogen atom to the surface and the HER activity would be low. Our calculations suggest that pure $Bi_2Te_3$ is not a good electrocatalyst.

As shown by the X-ray photoelectron spectroscopy (XPS) measurements, a $Bi_2Te_3$ surface exposed to air is usually not clean but oxidized[41,42]. In our experimental measurements, the acidity of the solution is very strong with pH=0.29 so the fully oxidized layers should dissolve into the solution. Only the partially oxidized layer may remain on the $Bi_2Te_3$ surface (see the XPS data analysis to be presented later). There may be a variety of oxidation structures. In the 2×2 surface unit cell of $Bi_2Te_3$ (001), we added one, two, and three oxygen atoms, which are labelled as 1O-, 2O-, and 3O-$Bi_2Te_3$. We have tried more than 20 oxidation structures as shown in Supplementary Fig. 5. To compare their stability, we calculated the formation energy of the oxidized $Bi_2Te_3$ slab as:

$$E_f = (E_{TI+nO} - E_{TI} - n \cdot E_O^{isol})/n, \quad (4)$$

where $E_{TI}$ is the energy of the clean $Bi_2Te_3$ slab, and $E_O^{isol}$ is the energy of an isolated O atom. $E_{TI+nO}$ is the total energy of the $n$O- $Bi_2Te_3$ slab, whose structure was obtained by optimizing the $n$O- $Bi_2Te_3$ slab adsorbed by one H atom and then removing this H atom without structural relaxation. The calculated $E_f$ values of various oxidation structures are showed in Fig. 5a. We found that when there is one O atom in one surface unit cell of $Bi_2Te_3$, the O atom prefers to form bonds with a Te and a Bi atom simultaneously. When one more O atom is present, the configuration, where the oxidized Te atom bonds with two O atoms, has a lower formation energy, and is labeled as 2O- $Bi_2Te_3$-a. When there are three O atoms in one surface unit cell, the oxidized

Te atom prefers to bond with three O atoms[42]. Among these structures with different oxidation levels, the 2O- $Bi_2Te_3$-a structure has the lowest formation energy.

We calculated the $\Delta G_H$ for the $Bi_2Te_3$ surfaces with different oxidation levels, and we found that when the H atom is bonded to the oxidized Te atom, $\Delta G_H$ is in the range of 0.20 and 0.35 eV, much smaller than that on the clean $Bi_2Te_3$ surface, indicating that the oxidation plays an important role in enhancing the HER activity. $Bi_2Te_3$ can be fully oxidized into $Bi_2O_3$ and $TeO_2$. We also calculated $\Delta G_H$ for the α-$Bi_2O_3$ (100)[43] and β-$TeO_2$ (001)[44] surfaces, as shown in Fig. 5b. Their $\Delta G_H$ are larger than 1.8 eV, so the fully oxidized $Bi_2Te_3$ is not a good HER catalyst. The high HER activity of the $Bi_2Te_3$ thin films is likely due to their partially oxidized surfaces. When the oxidized Te atom is bonded with one O atom, $\Delta G_H$ is larger than those for the 2O- and 3O- $Bi_2Te_3$ slabs. When the Te atom is oxidized by three O atoms, the structures may not be stable in the strong acidic solutions (see Fig. 5a). When the oxidized Te atom further forms bonds with four O atoms, the coordination number of the Te atom is the same as in the $TeO_2$ bulk. The 4O- $Bi_2Te_3$ slab has low HER activity and is not stable in the acidic environment.

It is likely that in the $Bi_2Te_3$ surfaces there are various oxidation structures. Considering the HER activity and the structural stability, the structure in which the oxidized Te atom is bonded with two O atoms should play an important role, so we proceed with the 2O-$Bi_2Te_3$-a configuration. Interestingly, when we switched off the spin–orbit coupling (SOC) interactions in our calculations, $\Delta G_H$ increases from 0.27 to 0.41 eV for the 2O-$Bi_2Te_3$-a surface, which means that the HER activity decreases. It is well known that without SOC, the TSSs of $Bi_2Te_3$ do not appear[10,45], so our calculations suggest that the TSSs of $Bi_2Te_3$ are of great importance to the HER activity of $Bi_2Te_3$.

In the acidic environment, the oxidized Te atom may be dissolved into the solution, which may create a Te vacancy. It has also been reported that the bulk of our MBE-grown $Bi_2Te_3$ thin films is an n-type semiconductor[46], which is likely attributed to Te vacancies[47,48]. We also calculated the $\Delta G_H$ for the $Bi_2Te_3$ surface with a Te vacancy by varying the H adsorption site. When the H atom is bonded with the Te atom that is in the same hexagonal ring where the removed Te atom is originally located (see Supplementary Fig. 6), $\Delta G_H$ is the lowest: 0.34 eV, which is also lower than $\Delta G_H$ for pure $Bi_2Te_3$ (Table 2). When we switched off SOC, $\Delta G_H$ increases to 0.48 eV, so the TSSs should also play an important role here.

Fig. 6 shows the band structures of various $Bi_2Te_3$ slabs before and after the H adsorption. It is clear that the Dirac cones move down after the H atom is adsorbed, indicating that the TSSs as electron baths accept electrons from the H atom. The similar Dirac cone shift can be also found when TIs work as substrates to tune the catalytic activity of overlaying materials[12,14]. We also calculated the local density of states (LDOS) on the absorbed H atom with and without SOC and the results are displayed in Supplementary Fig. 7. We found that after switching on SOC, the unoccupied states shift to higher energies, so the electron occupation of the H atomic orbitals decreases[12]. When the H atom loses electrons to the TSSs, $\Delta G_H$ becomes lower and it is easier for the H atom to adhere to the catalyst surface, so the electrocatalytic efficiency is enhanced. Note that in our calculations we only considered the flat basal surface of $Bi_2Te_3$, but in the experiments the spirally grown $Bi_2Te_3$ thin films have triangular domains, which would increase the surface-to-bulk ratio, and further enhance the electrocatalytic activity[40].

**XPS studies of $Bi_2Te_3$ thin films.** XPS is an effective technique for measuring the chemical states of the elements presented near the surface of a sample. Since our theoretical studies revealed that the high electrocatalytic performance of $Bi_2Te_3$ thin films may come from several specific oxidized structures, we have performed XPS depth-profiling together with detailed analysis on the 48 nm $Bi_2Te_3$ thin film to study the composition of oxides near its surface. Supplementary Fig. 8 displays the obtained XPS spectra near the Te *3d* (Supplementary Fig. 8a) and Bi *4f* (Supplementary Fig. 8b) core levels of the fresh surface and those after sputtered 10s, and 20s for this thin film. As can be seen in the top spectra in Supplementary Fig. 8a and Supplementary Fig. 8b, which were obtained before $Ar^+$ sputtering, a peak appears on the left hand side of each of the four Te and Bi core levels. These four peaks can be assigned to Te in $TeO_2$ and Bi in $Bi_2O_3$ because their binding energies are close to the reported standard values of 586.2 eV (Te $3d_{3/2}$), 575.9 eV (Te $3d_{5/2}$), 164.3 eV (Bi $4f_{5/2}$) and 159.0 eV (Bi $4f_{7/2}$) of these two oxides[49-51]. It is worthwhile to mention that, before sputtering, the two $TeO_2$ peaks are more pronounced than the two $Bi_2O_3$ peaks. After sputtering for 10s (middle spectra), the two peaks associated with $TeO_2$ are barely detectable, while the two peaks associated with $Bi_2O_3$ drop only slightly. These observations are consistent with the findings from Music *et al.*'s work[42] that $TeO_2$ dominates $Bi_2O_3$ at the surface because a $Bi_2Te_3$ unit-cell terminates with a top Te layer, however, at deeper regions down from the surface, $Bi_2O_3$ dominates over $TeO_2$ due to the fact that more Bi-O bonds are formed, which is attributed to the higher activity of Bi with oxygen than that of Te. After sputtering for 20s (bottom spectra), the two $TeO_2$ peaks become undetectable, while the two $Bi_2O_3$ peaks are barely detectable.

As can be seen in Supplementary Fig. 8a, the peaks of Te in $TeO_2$ are well separated from the peaks of Te in $Bi_2Te_3$. It is well known that the Te peaks of both $Bi_2Te_3$ and $TeO_2$ have symmetric shapes[52-55]. In the top spectrum of Fig. 7a, the raw data of the two Te $3d_{5/2}$ peaks with background subtracted are displayed together with two symmetric fitted curves. Obviously, an additional peak in the left shoulder of the peak of Te in $Bi_2Te_3$ must be added to obtain the best fit of the raw data as shown in the bottom spectrum. Similarly, an additional peak also must be added in Bi $4f_{7/2}$ to obtain the best fit, as shown in Fig. 7b. Based on this analysis, we have performed data fitting for the original XPS spectra by adding a peak at the left shoulder of each of the Bi or Te in $Bi_2Te_3$ peaks. Indeed, all the raw data are now well fitted as shown in Fig. 7c and d with all the binding energies of major peaks of $Bi_2Te_3$, $TeO_2$ and $Bi_2O_3$ consistent with the reported standard values[41,42,49-51]. Interestingly, previous XPS studies reported that an oxidized Te surface contains a small sub-oxide peak (TeO) besides the dominant $TeO_2$ peak[56,57], which locates also in the left shoulder of the Te core level, similar to what has been observed in our study. We believe, these additional peaks provide the evidence of the existence of the $Bi_2Te_3$ suboxide structures predicted from our theoretical studies. Since these additional peaks show higher intensity after the thin film was sputtered for 20s (as shown in the bottom spectra in Fig. 7 c and d) than that before sputtering, it is likely that these structures are mainly located below the two major oxides of $TeO_2$ and $Bi_2O_3$.

In the bottom spectra of Fig. 7a and b, the full width at half maximum (FWHM) values of all peaks are indicated. Interestingly, the FWHM of the suboxide peak of Te $3d_{5/2}$ (1.8eV) is twice that of pure $Bi_2Te_3$ (0.9 eV), while the FWHM of the suboxide peak of Bi $4f_{7/2}$(0.7eV) is the same as that of pure $Bi_2Te_3$ (0.7 eV). We believe, this observation can be linked to the differences in the bonding between Te and O and between Bi and O in the three most stable suboxide structures predicted by our theoretical studies. As shown in Supplementary Fig. 5, where the three marked specific suboxide structures (1O-$Bi_2O_3$, 2O-$Bi_2O_3$ and 3O-$Bi_2O_3$) are the most stable ones (our theoretical calculations reveal that this is true for either when hydrogen adsorption is taken into account or not), the Te atom linked to the O atom(s) is bonded with one, two or three O atom(s) respectively, while the corresponding Bi atom is always bonded with only one O atom in these three suboxide structures. Thus it is likely that the suboxide peak of Te $3d_{5/2}$ is attributed to the coexistence of multiple suboxide structures, as a consequence, it shows a much larger FWHM.

It is important to mention that the electrolyte used in our electrochemical measurements is a 0.5 M $H_2SO_4$ solution, which is able to etch away both the two major oxides of $TeO_2$ and $Bi_2O_3$ (together only 1-2 nm thick as estimated from the sputtering rate), such that the $Bi_2Te_3$ suboxide structures are exposed to act as the active catalytic material. Even if these two major oxides may not be completely removed, however our theoretical studies as described in the previous session show that their $\Delta G_H$ are larger than 1.8 eV, indicating that they will not contribute to the observed high HER performance of the $Bi_2Te_3$ thin films. Thus the three most stable suboxide structures are believed to play the roles, perhaps together with Te vacancies, in the HER activity.

**The role of TSSs for electrocatalytic activity.** Based on the above results, TI materials may in general offer high electrocatalytic performance. The role of TSSs can also be tested by replacing $Bi_2Te_3$ with another topological insulator material such as SnTe, which is a well-known topological crystalline insulator[58]. Supplementary Fig.9 displays the polarization curves of a 50 nm SnTe film and a Pt foil, indicating that SnTe also enjoys high electrocatalytic activity similar to $Bi_2Te_3$. Further experimental and theoretical study about the role of TSSs and oxidation of SnTe are underway.

# Conclusions

We have successfully synthesized high-quality $Bi_2Te_3$ thin films with different thicknesses on n+GaAs(111) substrates by the MBE technique. Their structural properties were examined by HRXRD, cross-sectional TEM and AFM, revealing that the spiral growth of the domains of the 48 nm $Bi_2Te_3$ thin film has been almost fully developed, which enjoys the maximum number of layers in each domain among these thin films. Their performances in HER were characterized by various electrochemical measurements. It was found that they all show high activity in HER, in particular, the 48 nm $Bi_2Te_3$ thin film exhibits the lowest overpotential, lowest Tafel slope and lowest charge–transfer resistance, which are among the best values achieved by existing candidates of non-noble metal catalysts. Our theoretical studies based on DFT calculations revealed an important finding that pure $Bi_2Te_3$ is not a good electrocatalyst; however, the $Bi_2Te_3$ thin films with partially oxidized surfaces, which are supported by the results of XPS studies, and Te vacancies can have high HER activity. We demonstrated that the TSSs play a key role in enhancing the electrocatalytic performance. The electrons in the H atoms transfer to the metallic surface states, which work as an effective electron bath, so the H atoms become easier to adhere to the $Bi_2Te_3$ surfaces; as a result, the HER performance is enhanced. The surface states protected by the time-reversal symmetry are resistant to non-magnetic impurities, so the HER performance would not be easily affected by contaminations, which is important for the durability of catalysts. Our study offers a new direction to optimize the performance of electrocatalytic materials without precious metals.

# Methods

**Materials synthesis.** All the $Bi_2Te_3$ samples studied in this work were fabricated on n+ GaAs (111)B substrates and the SnTe sample was prepared on n+ GaAs (100) substrates by a VG-V80H MBE system. For each sample of $Bi_2Te_3$, a 2.0-3.7nm ZnSe buffer layer was firstly deposited, followed by the growth of a $Bi_2Te_3$ layer with a desired thickness at substrate temperature of 242℃ (a lower temperature of 234℃ was used for the first 5 mins, it was found that this two-step growth mode provided a better structural quality and electrocatalytic activity) . All the growth processes were performed in an ultra-high-vacuum chamber with a basic pressure better than $1.0\times10^{-9}$ torr.

**Materials characterization.** All the $Bi_2Te_3$ samples were characterized by HRXRD (PANalytical Multipurpose X-Ray Diffractometer using Cu $K_{\alpha 1}$ x-rays with wavelength of 1.54056Å.), TEM (JEOL JEM-2010F with an acceleration voltage of 200 kV), and AFM (Dimension 3100 with a NanoScope IIIa controller (Digital Instruments) using tapping mode). All the XPS measurements were performed using Kratos-Axis Ultra DLD XPS *ex situ*. This instrument was equipped with a monochromatic Al Kα x-ray source (photon energy 1486.7 eV, 150 W) and the measurements were taken in hybrid lens mode with an energy step of 100 meV, a pass energy of 40 eV and a large measuring area of 1×2 $mm^2$. The ion sputtering of the film was handled using Ar ions with 4kV, 3×3 mm raster, 140 μA extractor current, and the sputtering rate was found to be similar to that of $SiO_2$, ~0.934 angstrom/s.

**Electrochemical measurements for HER.** All the electrochemical measurements were performed in a standard three-electrode electrolyzer connected to a CHI 660E electrochemical workstation (CH Instruments), using $Bi_2Te_3$/ZnSe/GaAs (111) B or SnTe/GaAs (001) as the working electrode, a Pt foil as a counter electrode, a saturated calomel electrode (SCE) served as the reference electrode, and 0.5 M $H_2SO_4$ as the electrolyte (sparged with $N_2$, purity ~99.99%). LSV was performed using a scan rate of 5mV $s^{-1}$. The ac impedance is measured with the frequency range of $10^6$ to 0.1 Hz with perturbation voltage amplitude of 5 mV. The impedance data were fitted to a simplified Randles circuit to extract the series and charge-transfer resistances. All data presented were *iR* corrected, where the solution resistances were determined by EIS experiments. The potential values shown were with respect to the reversible hydrogen electrode.

**Theoretical calculations.** We performed the calculations using the plane-wave DFT method as implemented in the Quantum Espresso code[59]. We used the Optimized Norm-Conserving Vanderbilt (ONCV) pseudopotentials[60-62] for the H, Bi, and Te atoms, and the ultrasoft pseudopotential[63] for the O atom. The kinetic energy cutoff is 60 Ry and the electron density cutoff is 480 Ry. The exchange-correlation functional is PBE[64]. The SOC is included in the self-consistent field (SCF) calculations. The lattice structure of $Bi_2Te_3$ is obtained from the Materials Project[65]. The $Bi_2Te_3$ slabs have three or four QLs terminated by Te atoms. The vacuum between two neighboring $Bi_2Te_3$ slabs is at least 15 Å. In the structural relaxation, we allowed the top 4 atomic layers to relax until the forces on atoms are smaller than 0.0001 Ry/bohr and fixed all the other atoms at their bulk positions. We used the 7 × 7 ×1 Monkhorst–Pack k-point mesh[66] for the 1 × 1 surface slab, and 3 × 3 × 1 for the 2 × 2 slab. In the SCF calculations, we used the 5 × 5 × 1 k-point mesh for the 2 × 2 slab.

We calculated the $\Delta G_H$ , as

$$\Delta G_H = \Delta E_H + \Delta E_{ZPE} - T\Delta S_{H.} \tag{5}$$

$\Delta E_H$ is the hydrogen adsorption energy calculated by:

$$\Delta E_H = E[Bi_2Te_3 + H^{ad}] - E[Bi_2Te_3] - \frac{1}{2}E[H_2], \tag{6}$$

where $E[Bi_2Te_3 + H^{ad}]$represents the total energy of a 3QL 2×2 surface slab of $Bi_2Te_3$ with one H atom adsorbed, $E[Bi_2Te_3]$ is the energy of the $Bi_2Te_3$ slab, and $E[H_2]$is the energy of a $H_2$ molecule in the gas phase. The change of the zero-point energy of a H atom, $\Delta E_{ZPE}$, is calculated by

$$\Delta E_{ZPE} = E_{ZPE}[H^{ad}] - \frac{1}{2}E_{ZPE}[H_2], \tag{7}$$

where $E_{ZPE}[H^{ad}]$represents the zero-point energy of one hydrogen atom adsorbed on $Bi_2Te_3$ surface, and $E_{ZPE}[H_2]$is the zero-point energy of a $H_2$ molecule in the gas phase. $\Delta S_H$is calculated as $-\frac{1}{2}S_{H_2}^0$, where $S_{H_2}^0$ is the entropy of $H_2$ in the gas phase at the standard condition ($130.68\ \mathrm{J} \cdot mol^{-1} \cdot K^{-1}\ at\ T = 298\ K\ and\ p = 1\ bar$)[67]. After combining the last two terms in equations (5), we can calculate $\Delta G_H$ simply as $\Delta G_H = \Delta E_H + 0.25\ eV$.

**Data availability**

The data that support the findings of this study are available from the authors on reasonable request; see author contributions for specific data sets.

# References

1 Seh, Z. W. *et al.* Combining theory and experiment in electrocatalysis: Insights into materials design. *Science* **355,** eaad4998 (2017).

2 Yin, H. *et al.* Ultrathin platinum nanowires grown on single-layered nickel hydroxide with high hydrogen evolution activity. *Nat. Commun.* **6,** 6430 (2015).

3 Yang, J. *et al.* Porous Molybdenum Phosphide Nano-Octahedrons Derived from Confined Phosphorization in UIO-66 for Efficient Hydrogen Evolution. *Angew. Chem. Int. Ed.* **55,** 12854-12858 (2016).

4 Peng, H. *et al.* Aharonov–Bohm interference in topological insulator nanoribbons. *Nat. Mater.* **9,** 225 (2009).

5 Fu, L., Kane, C. L. & Mele, E. J. Topological Insulators in Three Dimensions. *Phys. Rev. Lett.* **98,** 106803 (2007).

6 Chen, Y. L. *et al.* Single Dirac Cone Topological Surface State and Unusual Thermoelectric Property of Compounds from a New Topological Insulator Family. *Phys. Rev. Lett.* **105,** 266401 (2010).

7 Xu, Y., Gan, Z. & Zhang, S. C. Enhanced Thermoelectric Performance and Anomalous Seebeck Effects in Topological Insulators. *Phys. Rev. Lett.* **112,** 226801 (2014).

8 Alpichshev, Z. *et al.* STM imaging of electronic waves on the surface of $Bi_2Te_3$: topologically protected surface states and hexagonal warping effects. *Phys. Rev. Lett.* **104**, 016401 (2010).

9 Hsieh, D. *et al.* Observation of Unconventional Quantum Spin Textures in Topological Insulators. *Science* **323,** 919 (2009).

10 Hasan, M. Z. & Kane, C. L. Colloquium: topological insulators. *Reviews of modern physics* **82,** 3045 (2010).

11 Cha, J. J. & Cui, Y. The surface surfaces. *Nat. Nanotechnol.* **7,** 85 (2012).

12 Chen, H., Zhu, W., Xiao, D. & Zhang, Z. CO Oxidation Facilitated by Robust Surface States on Au-Covered Topological Insulators. *Phys. Rev. Lett.* **107,** 056804 (2011).

13 Xiao, J., Kou, L., Yam, C.-Y., Frauenheim, T. & Yan, B. Toward Rational Design of Catalysts Supported on a Topological Insulator Substrate. *ACS. Catal.* **5,** 7063-7067 (2015).

14 Li, L., Zeng, J., Qin, W., Cui, P. & Zhang, Z. Tuning the hydrogen activation reactivity on topological insulator heterostructures. *Nano Energy* **58,** 40-46 (2019).

15 He, Q. L., Lai, Y. H., Lu, Y., Law, K. T. & Sou, I. K. Surface Reactivity Enhancement on a Pd/$Bi_2Te_3$ Heterostructure through Robust Topological Surface States. *Sci. Rep.* **3,** 2497 (2013).

16 Rajamathi, C. R. *et al.* Weyl Semimetals as Hydrogen Evolution Catalysts. *Adv. Mater.* **29,** 1606202 (2017).

17 Rajamathi, C. R. *et al.* Photochemical Water Splitting by Bismuth Chalcogenide Topological Insulators. *ChemPhysChem* **18,** 2322-2327 (2017).

18 Li, G. *et al.* Carbon-Tailored Semimetal MoP as an Efficient Hydrogen Evolution Electrocatalyst in Both Alkaline and Acid Media. *Adv. Energy Mater.* **8,** 1801258 (2018).

19 Zhao, K., Ye, L., Tamargo, M. C. & Shen, A. Plasma-assisted MBE growth of ZnO on GaAs substrate with a ZnSe buffer layer. *Phys. Status Solidi C* **9,** 1809-1812 (2012).

20 Ramesh, S., Kobayashi, N. & Horikoshi, Y. High-quality ZnSe/GaAs superlattices: MEE growth, and structural and optical characterization. *J. Cryst. Growth* **111,** 752-756 (1991).

21 Heuken, M. *et al.* ZnS/ZnSe/GaAs heterostructures grown by metal-organic vapour phase epitaxy. *Mater. Sci. Eng. B* **9,** 189-193 (1991).

22 Blanton, T. N. *et al.* X-ray diffraction characterization of MOVPE ZnSe films deposited on (100) GaAs using conventional and high-resolution diffractometers. *Powder Diffr.* **24,** 78-81 (2009).

23 Kim, K.-C. *et al.* Free-electron creation at the 60° twin boundary in $Bi_2Te_3$. *Nat. Commun.* **7,** 12449 (2016).

24 Ginley, T. P. & Law, S. Growth of Bi2Se3 topological insulator films using a selenium cracker source. *J. Vac. Sci. Technol B* **34,** 02L105 (2016).

25 Li, H. D. *et al.* The van der Waals epitaxy of Bi2Se3on the vicinal Si(111) surface: an approach for preparing high-quality thin films of a topological insulator. *New J. Phys.* **12,** 103038 (2010).

26 Ambrosi, A., Sofer, Z., Luxa, J. & Pumera, M. Exfoliation of Layered Topological Insulators $Bi_2Se_3$ and $Bi_2Te_3$ via Electrochemistry. *ACS Nano* **10,** 11442-11448 (2016).

27 Kibsgaard, J., Chen, Z., Reinecke, B. N. & Jaramillo, T. F. Engineering the surface structure of $MoS_2$ to preferentially expose active edge sites for electrocatalysis. *Nat. Mater.* **11,** 963 (2012).

28 Yin, Y. *et al.* Contributions of Phase, Sulfur Vacancies, and Edges to the Hydrogen Evolution Reaction Catalytic Activity of Porous Molybdenum Disulfide Nanosheets. *JACS* **138,** 7965-7972 (2016).

29 Tsai, C. *et al.* Electrochemical generation of sulfur vacancies in the basal plane of $MoS_2$ for hydrogen evolution. *Nat. Commun.* **8,** 15113 (2017).

30 Yang, J., Wang, K., Zhu, J., Zhang, C. & Liu, T. Self-Templated Growth of Vertically Aligned 2H-1T $MoS_2$ for Efficient Electrocatalytic Hydrogen Evolution. *ACS. Appl. Mater. Inter.* **8,** 31702-31708 (2016).

31 Yang, J. *et al.* Integrated Quasiplane Heteronanostructures of $MoSe_2/Bi_2Se_3$ Hexagonal Nanosheets: Synergetic Electrocatalytic Water Splitting and Enhanced Supercapacitor Performance. *Adv. Funct. Mater.* **27,** 1703864 (2017).

32 Sharifi, T. *et al.* Thermoelectricity Enhanced Electrocatalysis. *Nano Lett.* **17,** 7908-7913 (2017).

33 Thomas, J. G. N. Kinetics of electrolytic hydrogen evolution and the adsorption of hydrogen by metals. *T. Faraday Soc.* **57,** 1603-1611 (1961).

34 Nørskov, J. K. *et al.* Trends in the exchange current for hydrogen evolution. *J. Electrochem. Soc.* **152,** J26 (2005).

35 Bose, R., Kim, T.-H., Koh, B., Jung, C. Y. & Yi, S. C. Influence of Phosphidation on $CoSe_2$ Catalyst for Hydrogen Evolution Reaction. *ChemistrySelect* **2,** 10661-10667 (2017).

36 Lee, C.P. *et al.* Beaded stream-like $CoSe_2$ nanoneedle array for efficient hydrogen evolution electrocatalysis. *J. Mater. Chem. A* **4,** 4553-4561 (2016).

37 Carim, A. I., Saadi, F. H., Soriaga, M. P. & Lewis, N. S. Electrocatalysis of the hydrogen-evolution reaction by electrodeposited amorphous cobalt selenide films. *J. Mater. Chem. A* **2,** 13835-13839 (2014).

38 Tang, Q. & Jiang, D.-e. Mechanism of Hydrogen Evolution Reaction on 1T-$MoS_2$ from First Principles. *ACS. Catal.* **6,** 4953-4961 (2016).

39 Zhang, H., Pan, Q., Sun, Z. & Cheng, C. Three-dimensional macroporous $W_2C$ inverse opal arrays for the efficient hydrogen evolution reaction. *Nanoscale* **11,** 11505-11512 (2019).

40 Zhu, J. *et al.* Boundary activated hydrogen evolution reaction on monolayer $MoS_2$. *Nat. Commun.* **10,** (2019).

41 Bando, H. *et al.* The time-dependent process of oxidation of the surface of $Bi_2Te_3$ studied by x-ray photoelectron spectroscopy. *J. Phys.: Condens. Matter* **12,** 5607-5616 (2000).

42 Music, D. *et al.* On atomic mechanisms governing the oxidation of $Bi_2Te_3$. *J. Phys.: Condens. Matter* **29,** 485705 (2017).

43 Lei, Y.-H. & Chen, Z.-X. Density functional study of the stability of various $\alpha$-$Bi_2O_3$ surfaces. *J. Chem. Phys* **138,** 054703 (2013).

44 Deringer, V. L., Stoffel, R. P. & Dronskowski, R. Thermochemical ranking and dynamic stability of $TeO_2$ polymorphs from ab initio theory. *Cryst. Growth Des* **14,** 871-878 (2014).

45 Qi, X.-L. & Zhang, S.-C. Topological insulators and superconductors. *Rev. Mod. Phys.* **83,** 1057 (2011).

46 He, H.-T. *et al.* Impurity Effect on Weak Antilocalization in the Topological Insulator $Bi_2Te_3$. *Phys. Rev. Lett.* **106,** 166805 (2011).

47 Chuang, P.-Y. *et al.* Anti-site defect effect on the electronic structure of a $Bi_2Te_3$ topological insulator. *RSC Adv.* **8,** 423-428 (2018).

48 Zhang, J. *et al.* Band structure engineering in $(Bi_{1-x}Sb_x)_2Te_3$ ternary topological insulators. *Nat. Commun.* **2,** 574 (2011).

49 Dharmadhikari, V. S., Sainkar, S. R., Badrinarayan, S. & Goswami, A. Characterisation of thin films of bismuth oxide by X-ray photoelectron spectroscopy. *J. Electron. Spectrosc. Relat. Phenom.* **25,** 181-189 (1982).

50 Escobar-Alarcón, L. *et al.* Preparation and characterization of bismuth nanostructures deposited by pulsed laser ablation. *J. Phys. Conf. Ser.* **582,** 012013 (2015).

51 Wu, M. *et al.* Phosphine-free engineering toward the synthesis of metal telluride nanocrystals: the role of a Te precursor coordinated at room temperature. *Nanoscale* **10,** 21928-21935 (2018).

52 Mekki, A., Khattak, G. D. & Wenger, L. E. Structural and magnetic investigations of $Fe_2O_3$–$TeO_2$ glasses. *J. Non-Cryst. Solids* **352,** 3326-3331 (2006).

53 Mekki, A., Khattak, G. D. & Wenger, L. E. Structural and magnetic properties of $MoO_3$–$TeO_2$ glasses. *J. Non-Cryst. Solids* **351,** 2493-2500 (2005).

54 Hoefer, K. *et al.* Intrinsic conduction through topological surface states of insulating $Bi_2Te_3$ epitaxial thin films. *PNAS* **111,** 14979 (2014).

55 Fornari, C. I. *et al.* Preservation of pristine $Bi_2Te_3$ thin film topological insulator surface after ex situ mechanical removal of Te capping layer. *APL Mater.* **4,** 106107 (2016).

56 Berchenko, N. *et al.* Surface oxidation of SnTe topological crystalline insulator. *Appl. Surf. Sci.* **452,** 134-140 (2018).

57 Kosmala, T. *et al.* Metallic Twin Boundaries Boost the Hydrogen Evolution Reaction on the Basal Plane of Molybdenum Selenotellurides. *Adv. Energy Mater.* **8,** 1800031 (2018).

58 Hsieh, T. H. *et al.* Topological crystalline insulators in the SnTe material class. *Nat. Commun.* **3,** 982 (2012).

59 Giannozzi, P. *et al.* Advanced capabilities for materials modelling with Quantum ESPRESSO. *J. Phys.: Condens. Matter* **29,** 465901 (2017).

60 Hamann, D. R. Optimized norm-conserving Vanderbilt pseudopotentials. *Phys. Rev. B* **88,** (2013).

61 Schlipf, M. & Gygi, F. Optimization algorithm for the generation of ONCV pseudopotentials. *Comput. Phys. Commun.* **196,** 36-44 (2015).

62 Scherpelz, P., Govoni, M., Hamada, I. & Galli, G. Implementation and validation of fully relativistic GW calculations: spin–orbit coupling in molecules, nanocrystals, and solids. *J. Chem.Theory Comput* **12,** 3523-3544 (2016).

63 Vanderbilt, D. Soft self-consistent pseudopotentials in a generalized eigenvalue formalism. *Phys. Rev. B* **41,** 7892 (1990).

64 Perdew, J. P., Burke, K. & Ernzerhof, M. Generalized gradient approximation made simple. *Phys. Rev. Lett.* **77,** 3865 (1996).

65 Jain, A. *et al.* Commentary: The Materials Project: A materials genome approach to accelerating materials innovation. *Apl Mater.* **1,** 011002 (2013).

66 Monkhorst, H. J. & Pack, J. D. Special points for Brillouin-zone integrations. *Phys. Rev.B* **13,** 5188-5192 (1976).

67 Atkins, P. W. & De Paula, J. *Atkins' Physical chemistry*. (Oxford Univ. Press, 2006).

# Acknowledgement

We thank Junwei Liu, Zhipan Liu and Dingyong Zhong for their helpful discussions. This research was funded by the Research Grants Council of the Hong Kong Special Administrative Region, China, grant number 16304515 and William Mong Institute of Nano Science and Technology, project number WMINST19SC07. D.P. acknowledges support from the Croucher Foundation through the Croucher Innovation Award and the Energy Institute at the Hong Kong University of Science and Technology.

# Author Contributions

Q.Q. and I.K.S initiated this study and further designed the experiments; Q.Q. and J.L. carried out the sample synthesis; Q.Q. conducted the electrochemical measurements and characterizations; B.L. and D.P. carried out the theoretical calculations; Q.Q., I.K.S., D.P., and B.L. wrote the manuscript. All authors performed the data analysis and discussions.

# Additional information

**Supplementary Information** accompanies this paper as a single PDF.

**Competing interests:** The authors declare no competing interests.

# Figures

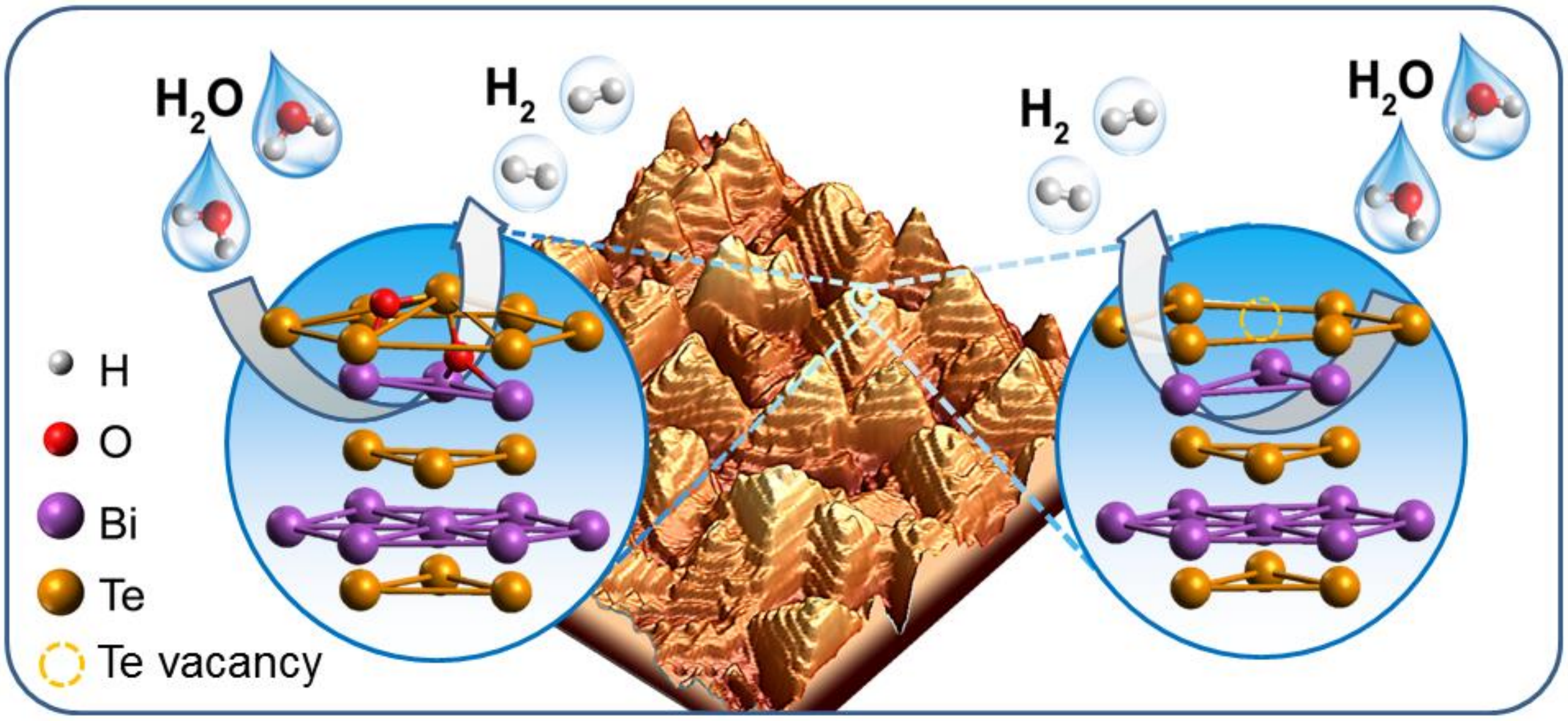

**Fig. 1 | Schematic diagram.** A $Bi_2Te_3$ thin film with partially oxidized surfaces or Te vacancies exhibits highly efficient hydrogen evolution reaction (HER) activity.

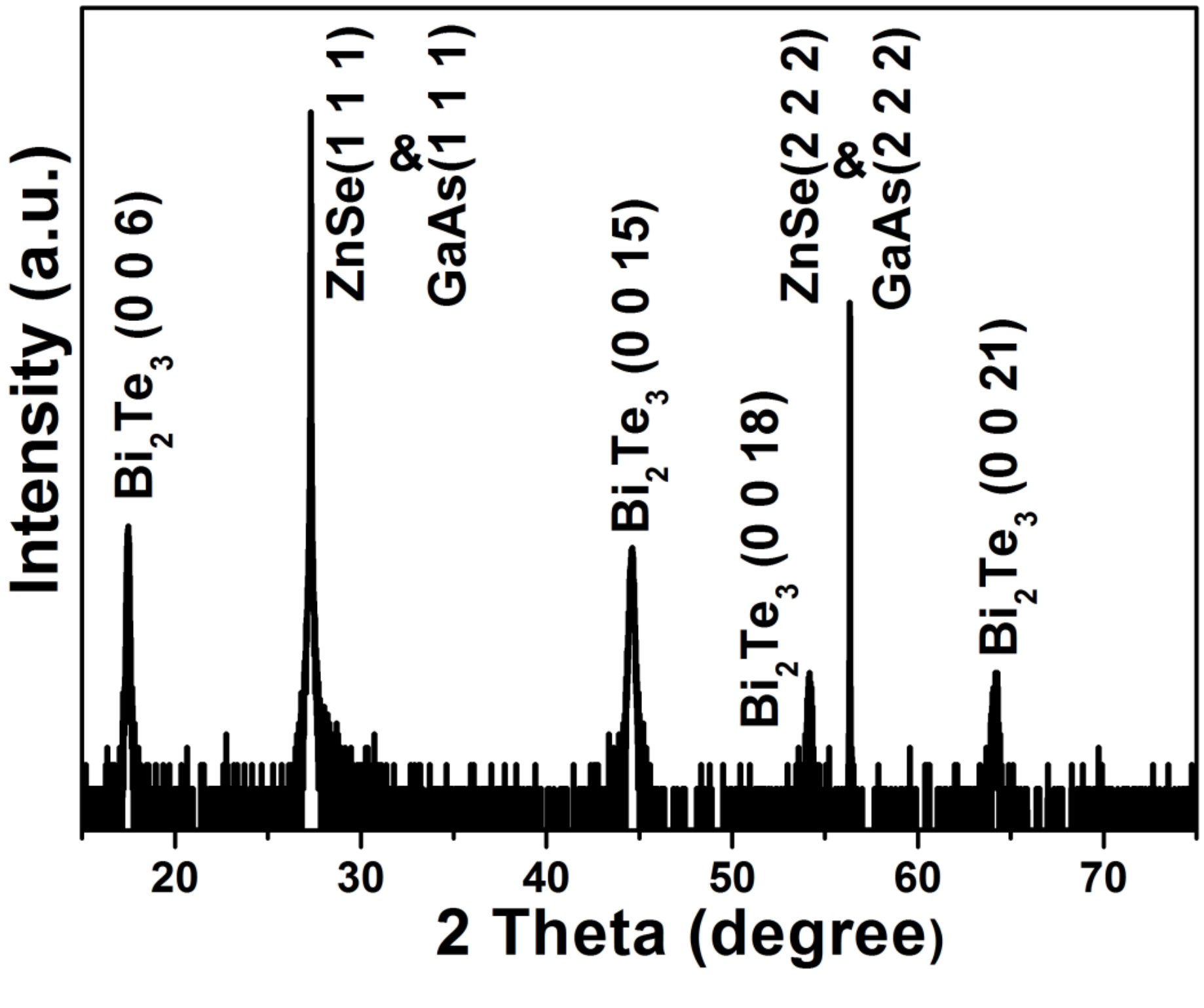

**Fig. 2 | High resolution X-ray diffraction (HRXRD) profile of a typical $Bi_2Te_3$ sample.**

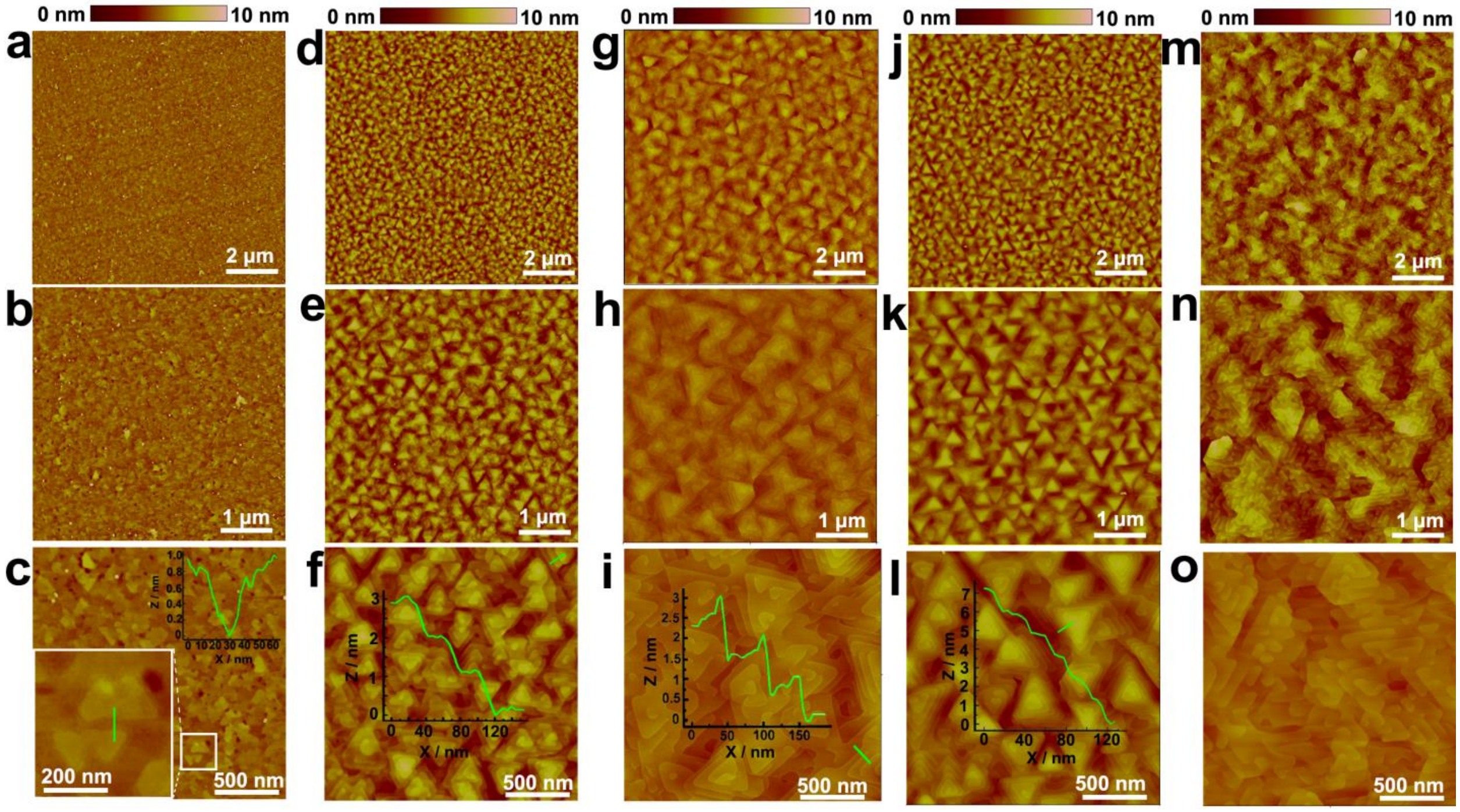

**Fig. 3 | Morphology analysis of $Bi_2Te_3$ thin films.** Atomic force microscopy (AFM) images of $Bi_2Te_3$ films with nominal thicknesses of (**a**)-(**c**) 5 nm, (**d**)-(**f**) 17 nm, (**g**)-(**i**) 34 nm, (**j**)-(**l**) 48 nm and (**m**)-(**o**) 61 nm.

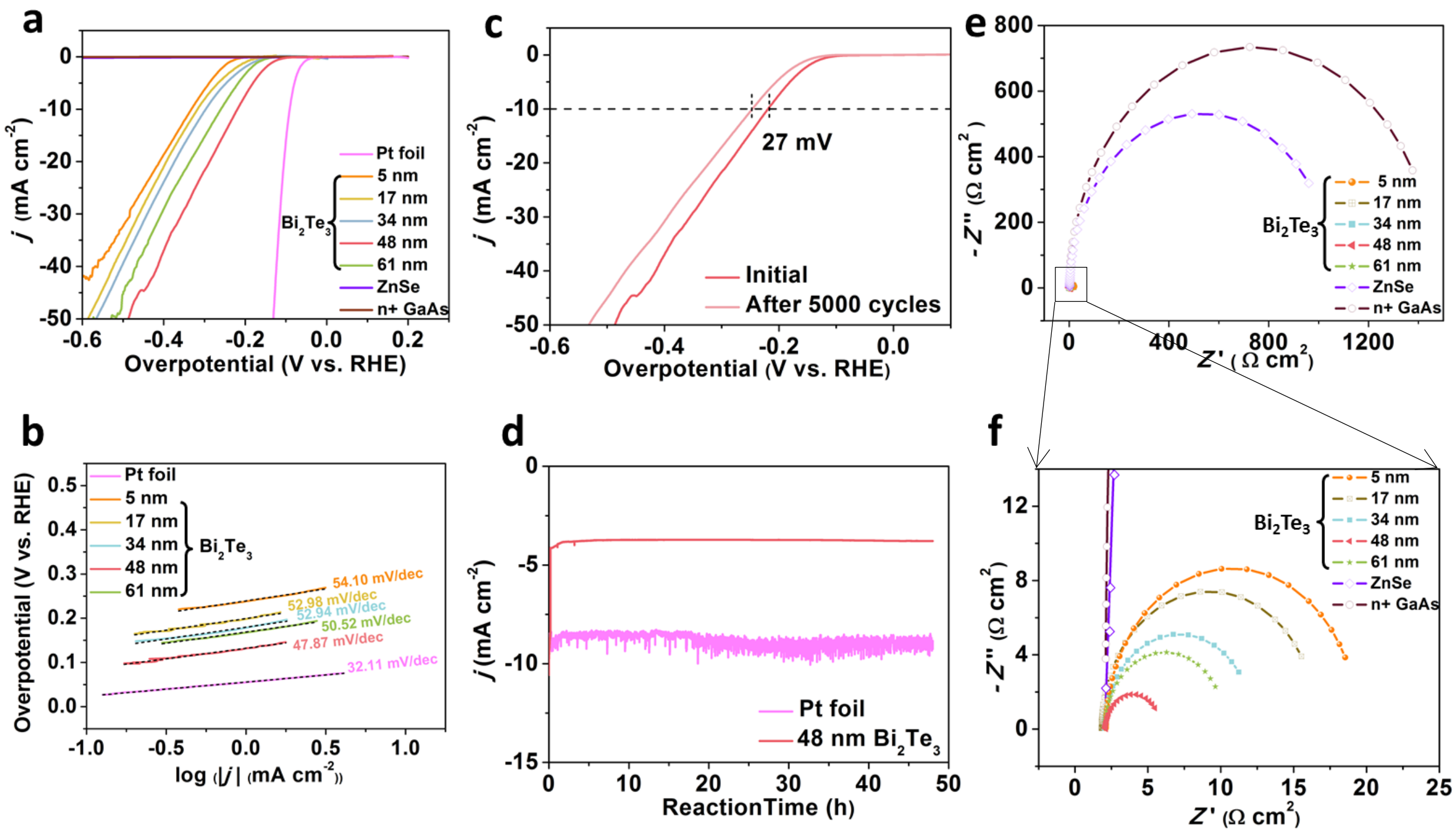

**Fig. 4 | HER electrocatalytic performances of molecular beam epitaxy (MBE)-grown $Bi_2Te_3$ thin films.** **a** Polarization curves (*iR*-corrected) of as–grown $Bi_2Te_3$ with different thicknesses, ZnSe buffer, n+GaAs substrate and commercial Pt foil. **b** Corresponding Tafel plots of the active materials in (**a**). **c** Polarization curves (*iR*-corrected) of the 48 nm $Bi_2Te_3$ film recorded before and after 5000 cycles of cyclic voltammetry(CV) using accelerated degradation tests (scan rate: 100 mV $s^{-1}$). **d** Chronoamperometry (CP) test results of the 48 nm $Bi_2Te_3$ thin film and commercial Pt foil at current density($j$)=10mA $cm^{-2}$ respectively. **e, f** Electrochemical impedance spectroscopy (EIS) of the $Bi_2Te_3$ thin films with different thicknesses, ZnSe buffer and n+ GaAs substrate.

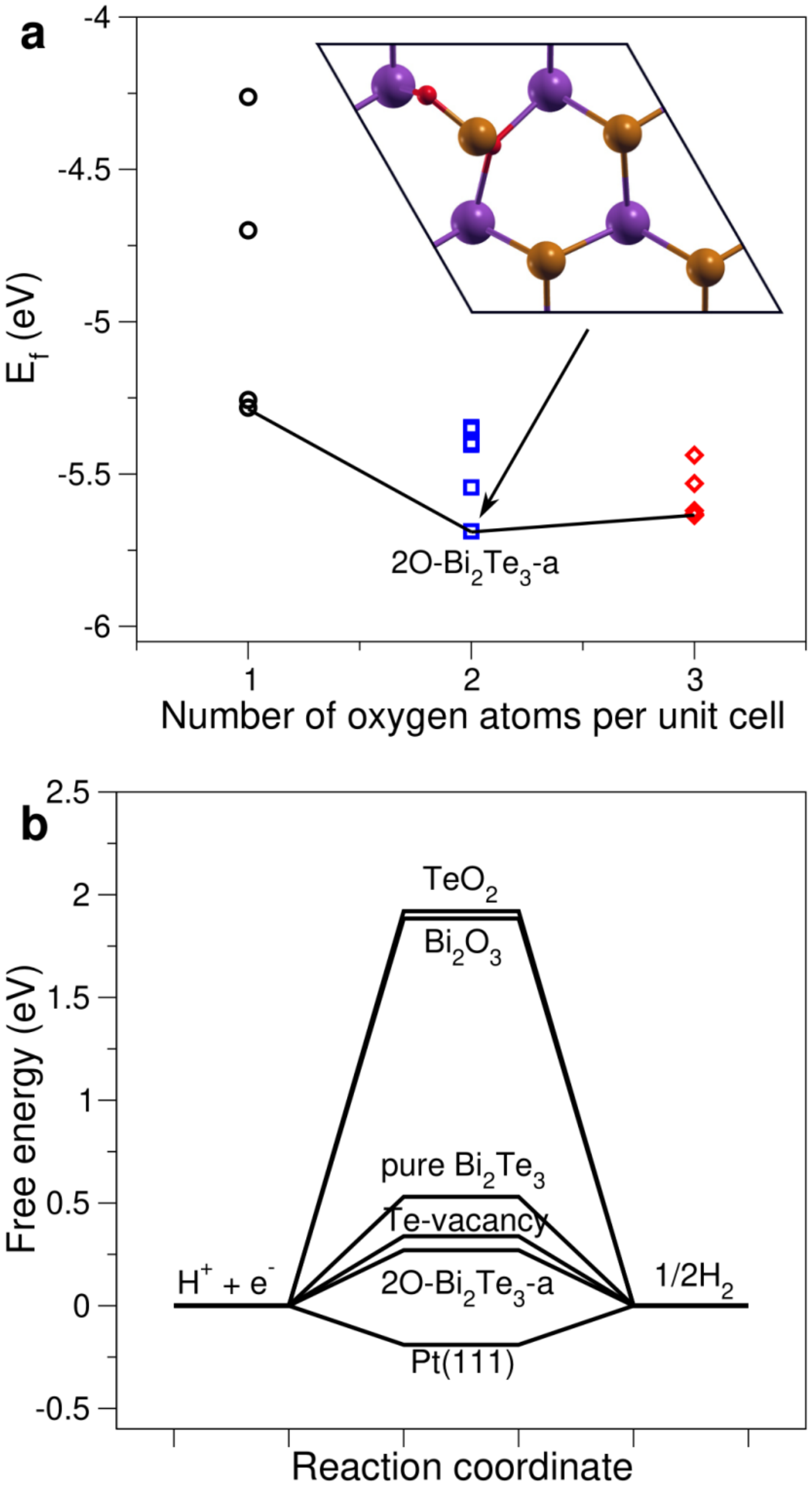

**Fig. 5 | Formation energies ($E_f$) of the partially oxidized $Bi_2Te_3$ structures and the free energy of hydrogen adsorption ($\Delta G_H$). a** The $E_f$ of oxidized $Bi_2Te_3$ slabs at various oxidation levels. The inserted figure shows the top view of the top two atomic layers of the oxidized $Bi_2Te_3$ (001) surface: 2O-$Bi_2Te_3$-a. **b** The $\Delta G_H$ on the β-$TeO_2$ (001) surface, the α-$Bi_2O_3$ (100) surface, the pure, a Te-vacancy-containing and the partially oxidized (2O-$Bi_2Te_3$-a) $Bi_2Te_3$ (001) surfaces, and the Pt (111) surface.

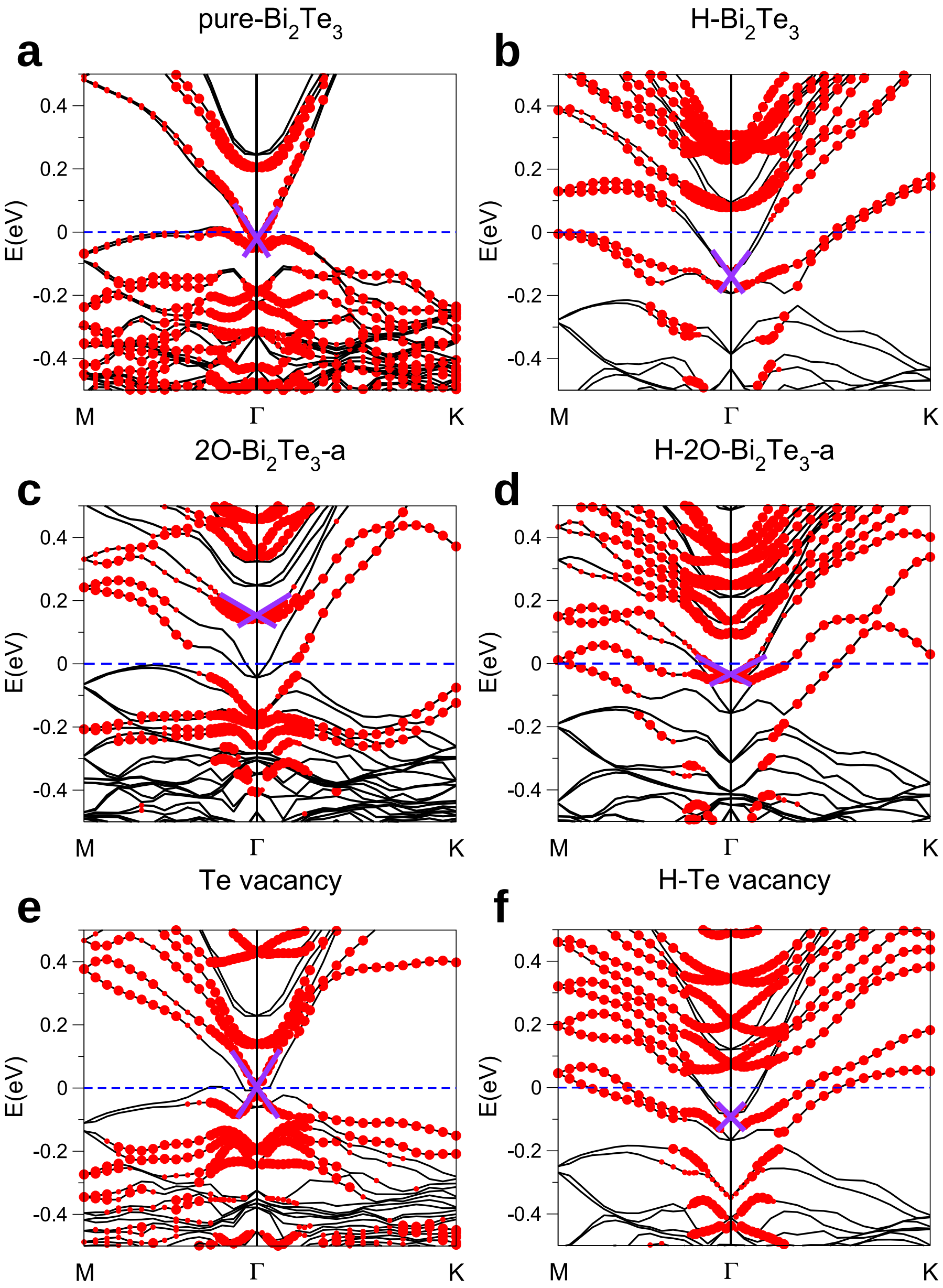

**Fig. 6 | Band structures of various $Bi_2Te_3$ basal slabs before (left panel) and after (right panel) hydrogen adsorption. (a)** and (**b**) pure $Bi_2Te_3$, (**c**) and (**d**) 2O-$Bi_2Te_3$-a structure, and (**e**) and (**f**) $Bi_2Te_3$ with a Te vacancy. The sizes of the red dots represent the contributions from the upper-surface quintuple layers (QL) of $Bi_2Te_3$. The purple crosses indicate the upper-surface TSSs. In (**a**), (**b**), (**e**) and (**f**), the $Bi_2Te_3$ slabs have 3 QLs, and in (**c**) and (**d**), the slabs have 4 QLs. The Fermi levels of the slabs are set to zero.

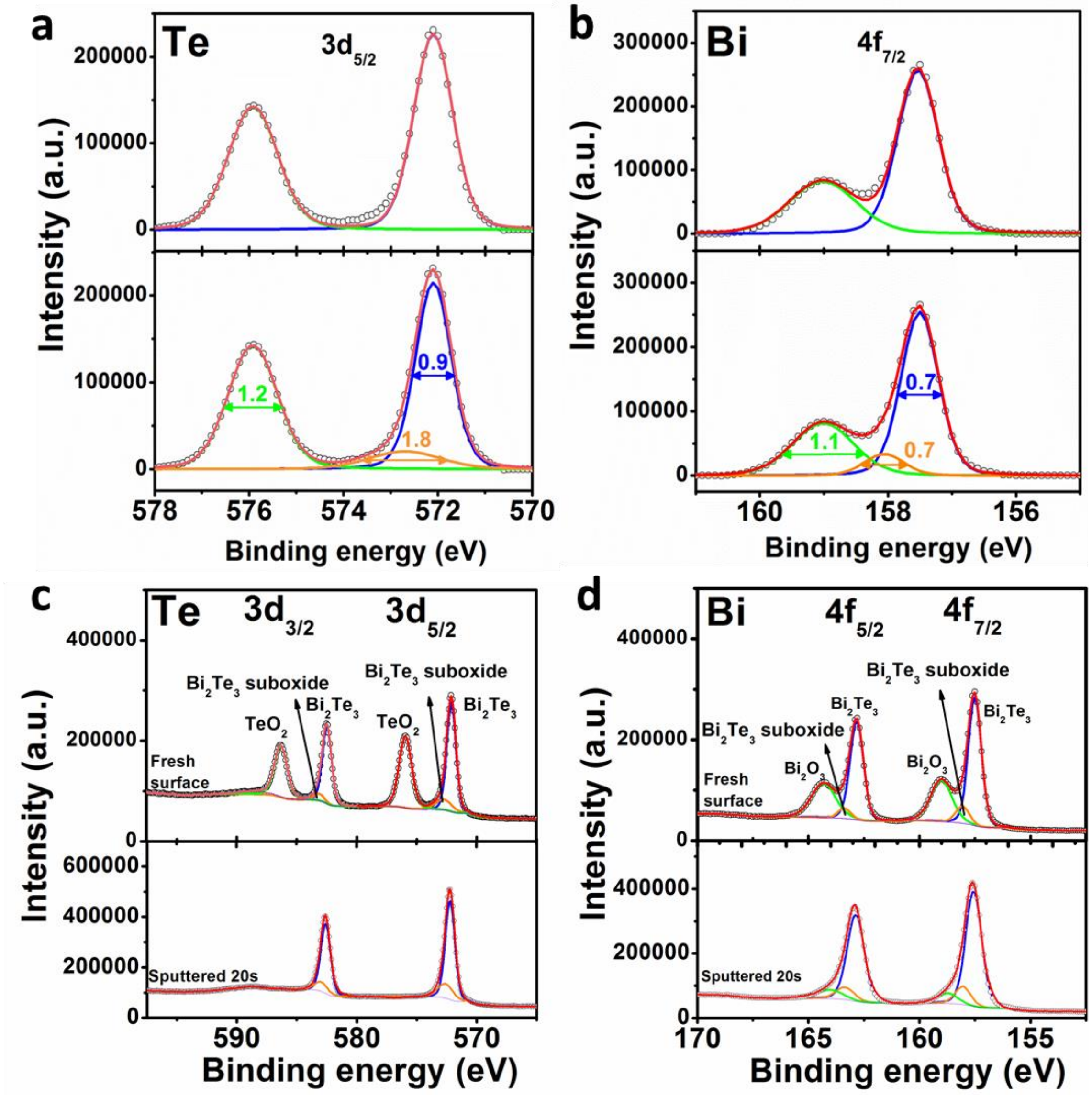

**Fig. 7 | X-ray photoelectron spectroscopy (XPS) spectra obtained from the surface of the 48 nm $Bi_2Te_3$ thin film exposed to air after the MBE growth.** **a** Te *3d*$_{5/2}$ and **b** Bi *4f*$_{7/2}$ with Shirley background subtraction, the top and bottom graphs also display the fitting curves before and after adding an additional peak of $Bi_2Te_3$ suboxide structures respectively. **c** Te *3d* and **d** Bi *4f*, the top and bottom graphs show the spectra of the fresh surface and those after sputtered for 20s respectively, together with the fitted curves by including $Bi_2Te_3$ suboxide structures.

# Tables

**Table 1 | Comparison of the HER performances between $Bi_2Te_3$ films (48 nm) and other nanomaterials, including $Bi_2Se_3$, $Sb_2Te_3$, $Bi_{0.5}Sb_{1.5}Te_3$ nanosheets, exfoliated $Bi_2Te_3$ nanosheets and some molybdenum dichalcogenide($MX_2$)-based catalysts. The electrolytes used for obtaining all these data are the same (0.5M $H_2SO_4$).**

| Catalyst | Overpotential ($\eta$) (mV vs. RHE) for $j$ =-10mA $cm^{-2}$ | Tafel slope (mV/ dec) |
|---|---|---|
| $Bi_2Te_3$ 48 nm thin film (this work) | 219 | 47.87 |
| $Bi_2Se_3$ nanosheets [31] | ~500 | 78 |
| $MoSe_2$ nanosheets [31] | ~300 | 100 |
| $Bi_2Se_3$+$MoSe_2$ nanosheets mixture [31] | ~340 | 98 |
| $MoSe_2$/$Bi_2Se_3$ nanostructured hybrids [31] | ~250 | 44 |
| $Sb_2Te_3$ nanosheets [32] | ~510 | N/A |
| $Bi_{0.5}Sb_{1.5}Te_3$ nanosheets [32] | ~520 | N/A |
| $Bi_{0.5}Sb_{1.5}Te_3$ nanosheets on carbon paper (temperature gradient=90 °C) [32] | ~400 | N/A |
| Exfoliated $Bi_2Te_3$ nanosheets [26] | 680 | N/A |
| Double-gyroid $MoS_2$ (Mo electrodeposition time: 1 min) [27] | 285 | 50 |
| Vertically aligned homostructured $MoS_2$ nanosheets [30] | 202 | 60 |
| Fewer layer 1T phase $MoS_2$ [30] | 276 | 69 |
| 2H phase $MoS_2$ nanosheets [28] | 343 | 106 |
| 1T phase $MoS_2$ nanosheets [28] | 203 | 48 |
| Monolayer $MoS_2$ after desulfurization [29] | 320 | 102 |

**Table 2 | The free energy of hydrogen adsorption with ($\Delta G_{H,SOC\ on}$) and without ($\Delta G_{H,SOC\ off}$) the spin-orbital coupling (SOC) on the pure, the partially oxidized (2O-$Bi_2Te_3$-a) and the Te-vacancy-containing $Bi_2Te_3$ surfaces. $\Delta G_{SOC} = \Delta G_{H,SOC\ off} - \Delta G_{H,SOC\ on}$. (Unit: eV)**

| structure | $\Delta G_{H,SOC\ off}$ | $\Delta G_{H,SOC\ on}$ | $\Delta G_{SOC}$ |
|---|---|---|---|
| Pure $Bi_2Te_3$ | 0.718 | 0.533 | 0.185 |
| 2O-$Bi_2Te_3$-a | 0.411 | 0.274 | 0.137 |
| Te-vacancy-containing $Bi_2Te_3$ | 0.478 | 0.342 | 0.136 |

# Expediting hydrogen evolution through topological surface states on $Bi_2Te_3$

**Qing Qu[1,3†], Bin Liu[2†], Jing Liang[2,3], Ding Pan[2,4]* and Iam Keong Sou[1,2,3]***

**[1]Nano Science and Technology Program, The Hong Kong University of Science and Technology, Hong Kong, China.**

**[2]Department of Physics, The Hong Kong University of Science and Technology, Clear Water Bay, Hong Kong , China.**

**[3]William Mong Institute of Nano Science and Technology, The Hong Kong University of Science and Technology, Hong Kong, China**

**[4]Department of Chemistry, The Hong Kong University of Science and Technology, Hong Kong, China.**

**[†] These authors contributed equally to this work.**

***Corresponding authors**

Supplementary Information for ***Nature Communications***

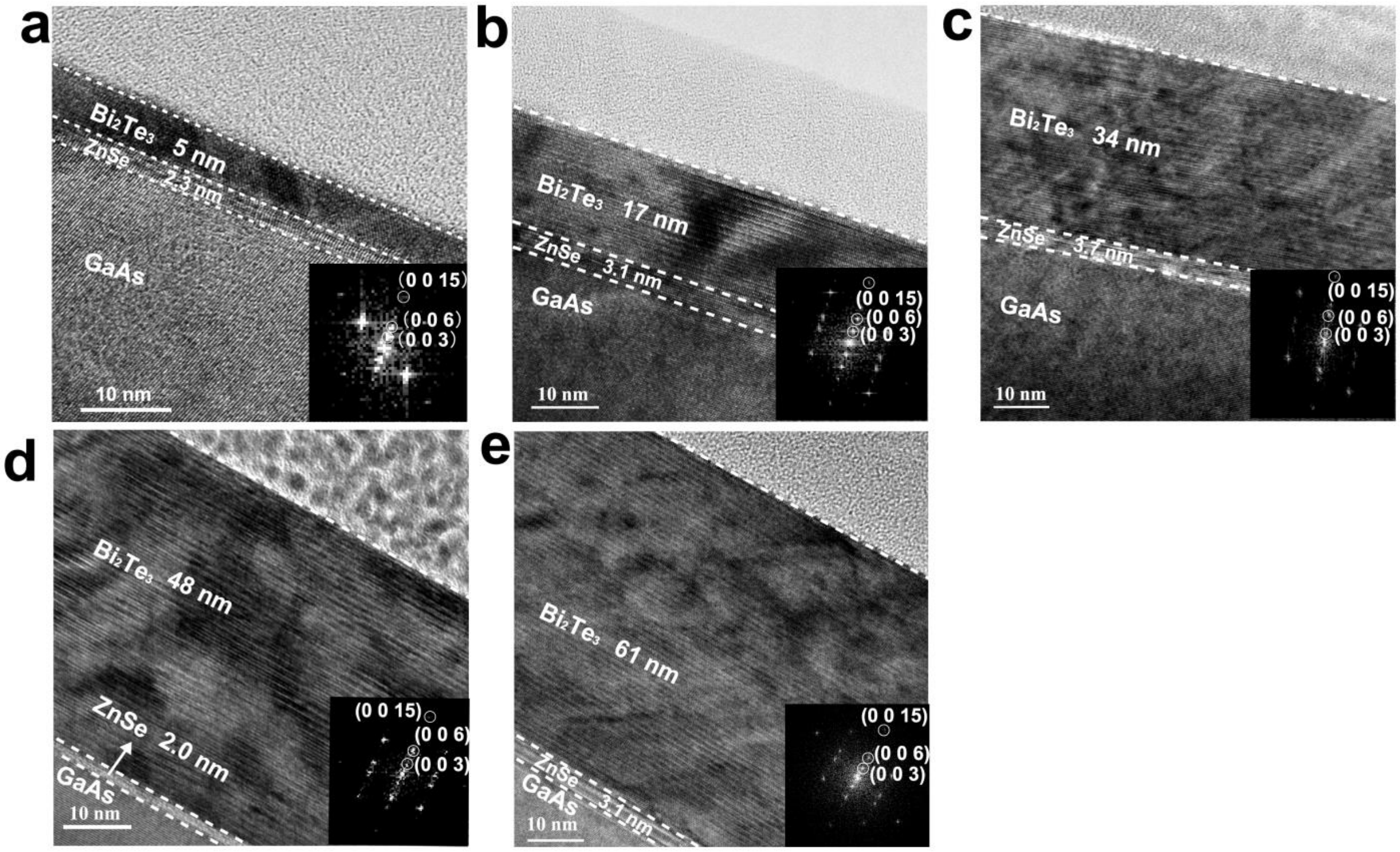

**Supplementary Fig.1|** Cross-sectional transmission electron microscopy (TEM) images of $Bi_2Te_3$ thin films. with thicknesses of (a) 5 nm, (b) 17 nm, (c) 34 nm, (d) 48 nm and (e) 61 nm. The insets are fast Fourier transform (FFT) patterns of $Bi_2Te_3$.

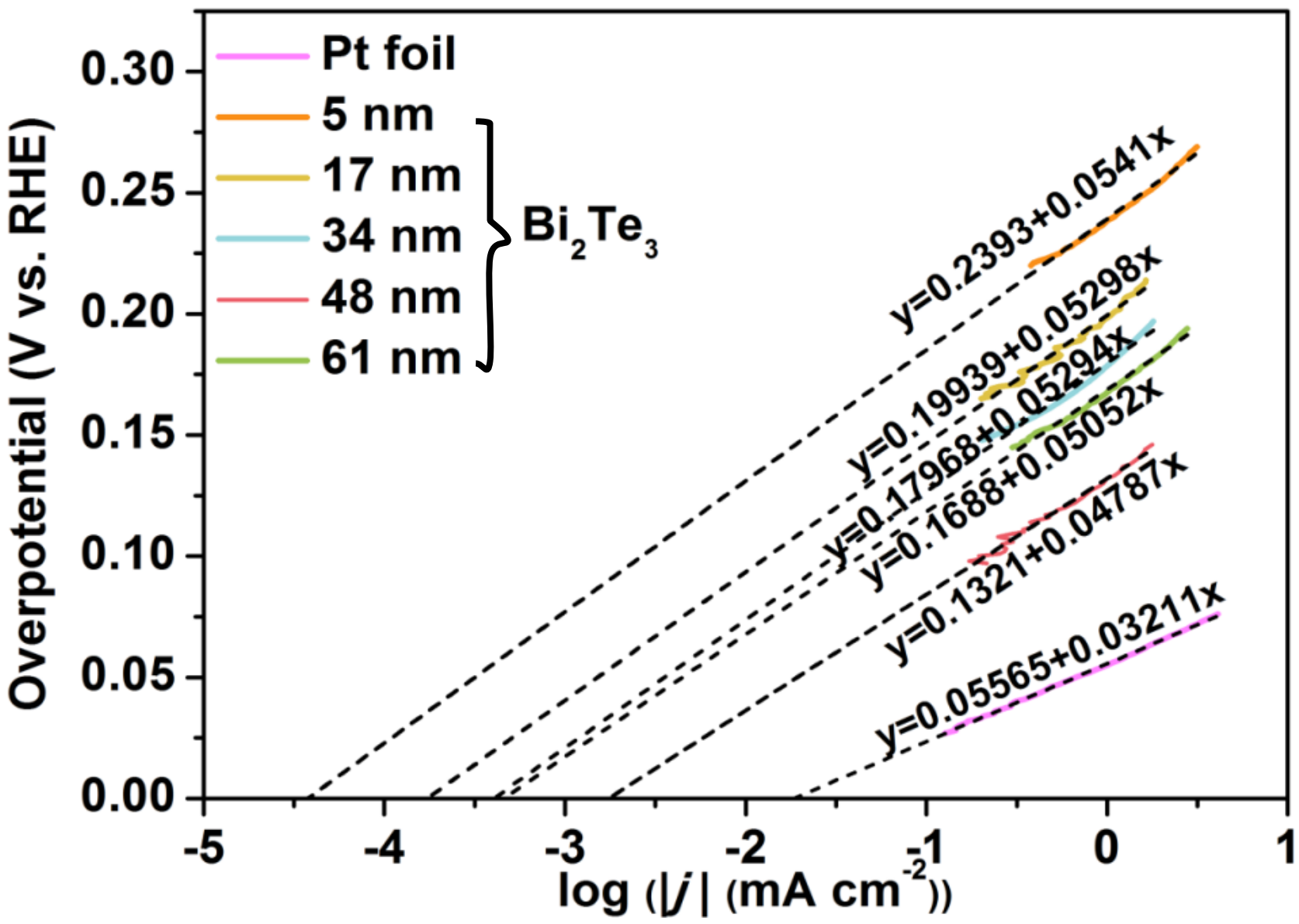

**Supplementary Fig.2 |** Exchange current densities ($j_0$) for the active materials in Fig.3a, derived from the Tafel plots in Fig. 3b, as indicated by the x-intercepts of the dash lines.

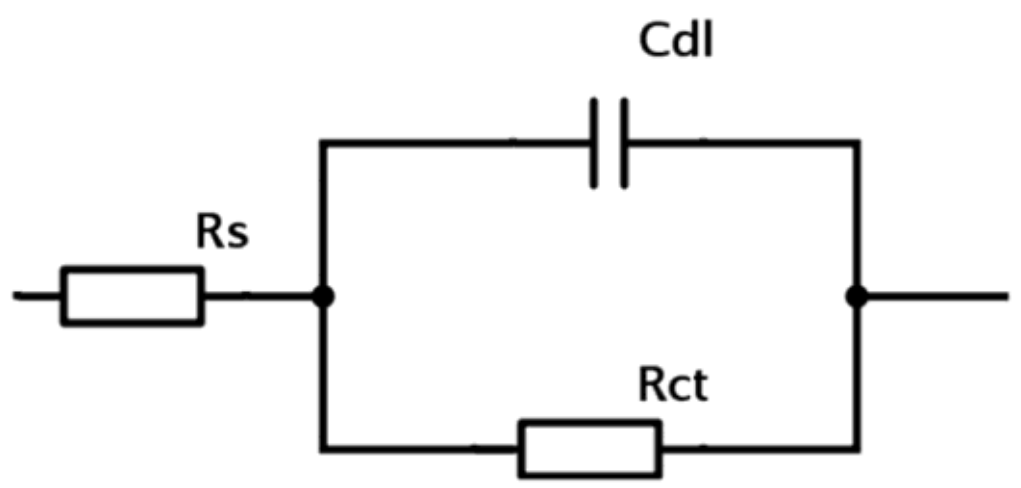

**Supplementary Fig. 3 |** Simplified Randles circuit. Equivalent circuit models used for fitting the EIS response of HER, where $R_s$ is the electrolyte resistance and $R_{ct}$ denote the charge-transfer resistance.

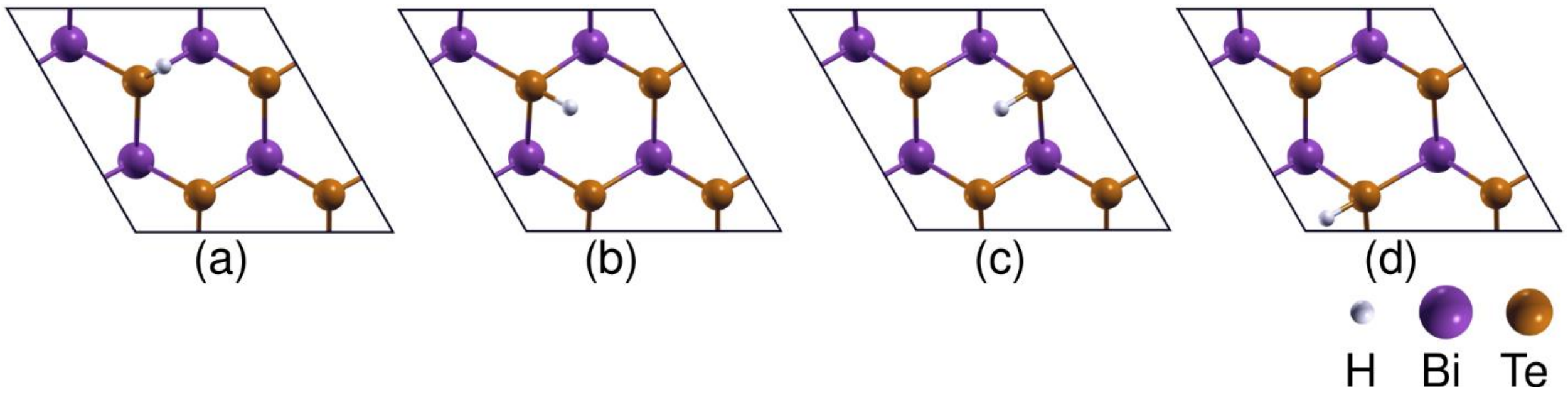

**Supplementary Fig. 4 |** Structures of a pure $Bi_2Te_3$ (001) surface adsorbed by a H atom with different initial adsorption positions. Only top two atomic layers are showed in the top view.

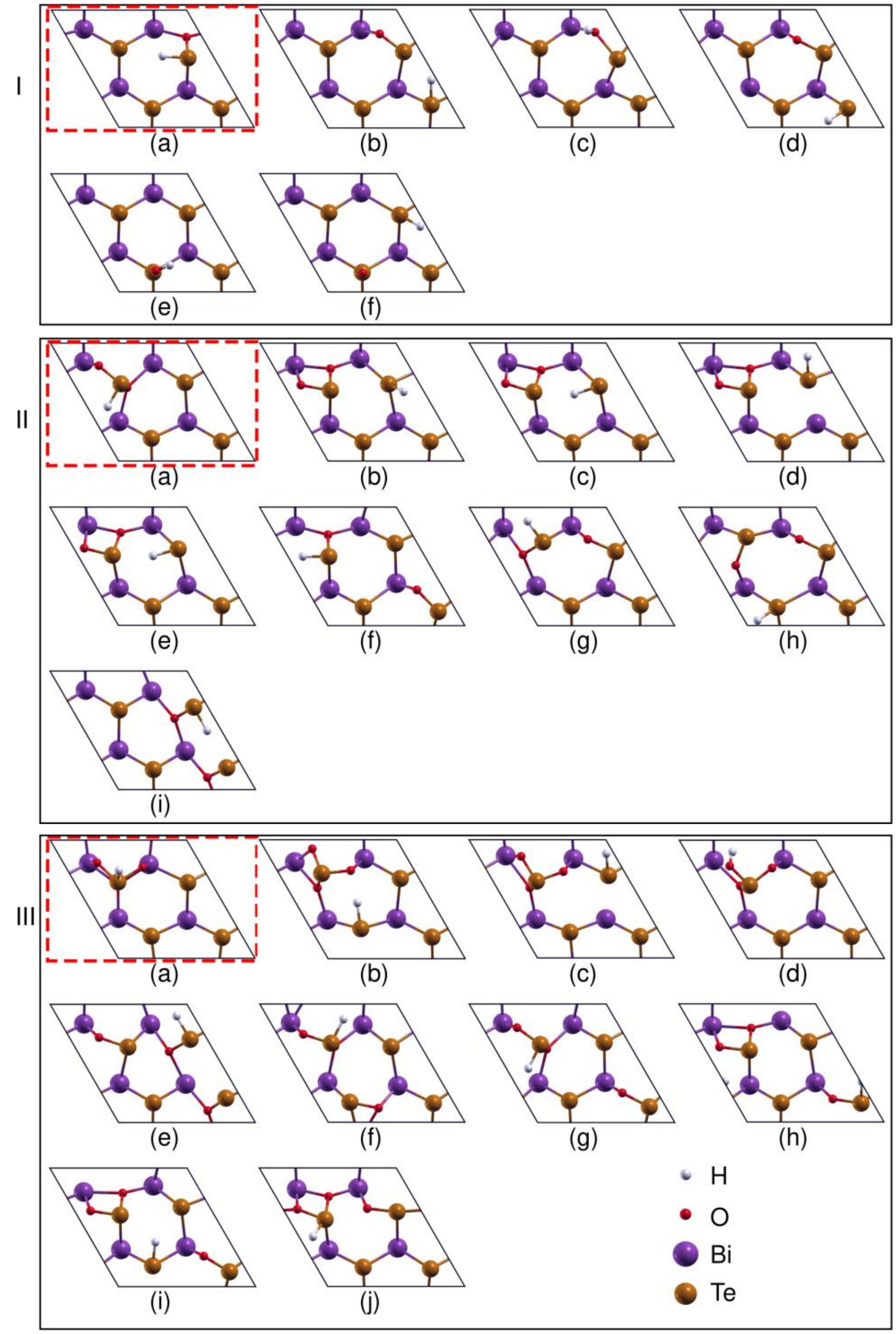

**Supplementary Fig. 5 |** Structures of (I) 1O-, (II) 2O-, and (III) 3O- $Bi_2Te_3$ (001) surfaces adsorbed by a H atom with different initial adsorption positions. Only top two atomic layers are showed in the top view. The most energetically favored structures are labelled by dashed boxes.

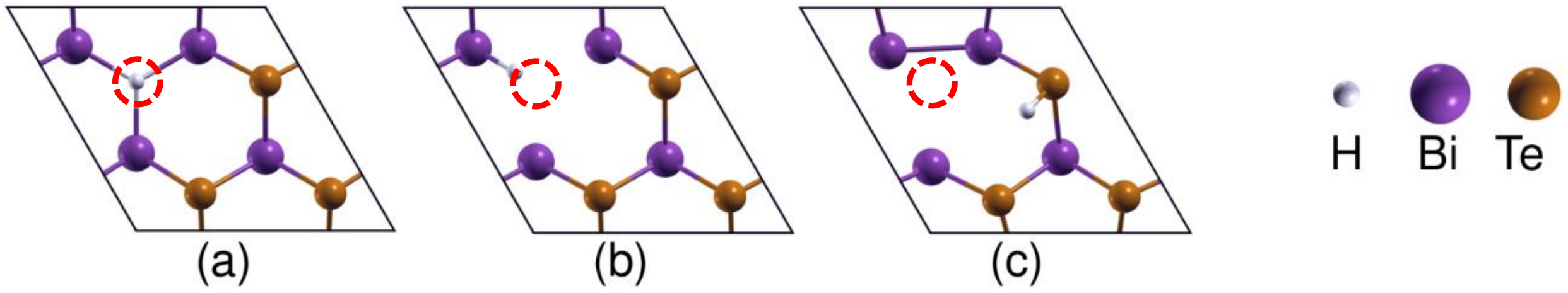

**Supplementary Fig. 6 |** Structures of the Te-vacancy-containing $Bi_2Te_3$ (001) surface adsorbed by a H atom with different initial adsorption positions. Only top two atomic layers are showed in the top view. The red dashed circle indicates the position of the Te vacancy. Structure (c) has the lowest $\Delta G_H$.

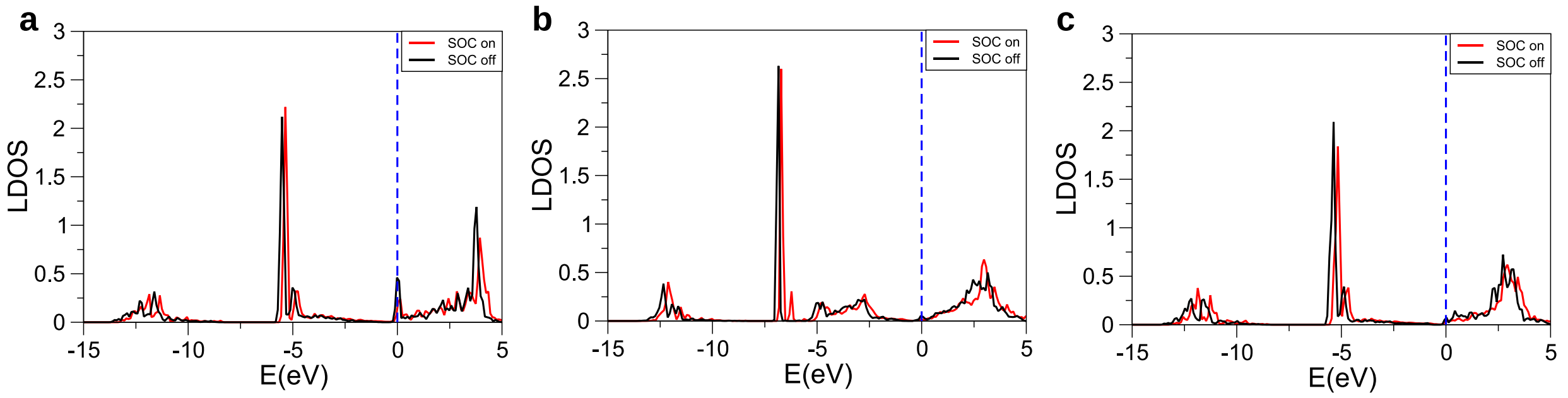

**Supplementary Fig. 7 |** Calculated local density of states (LDOS) of the H atom absorbed on pure (a) $Bi_2Te_3$, (b) 2O-$Bi_2Te_3$-a and (c) Te-vacancy-containing $Bi_2Te_3$ slabs with (red line) and without (black line) SOC interactions. The Fermi levels of the slabs are set to zero.

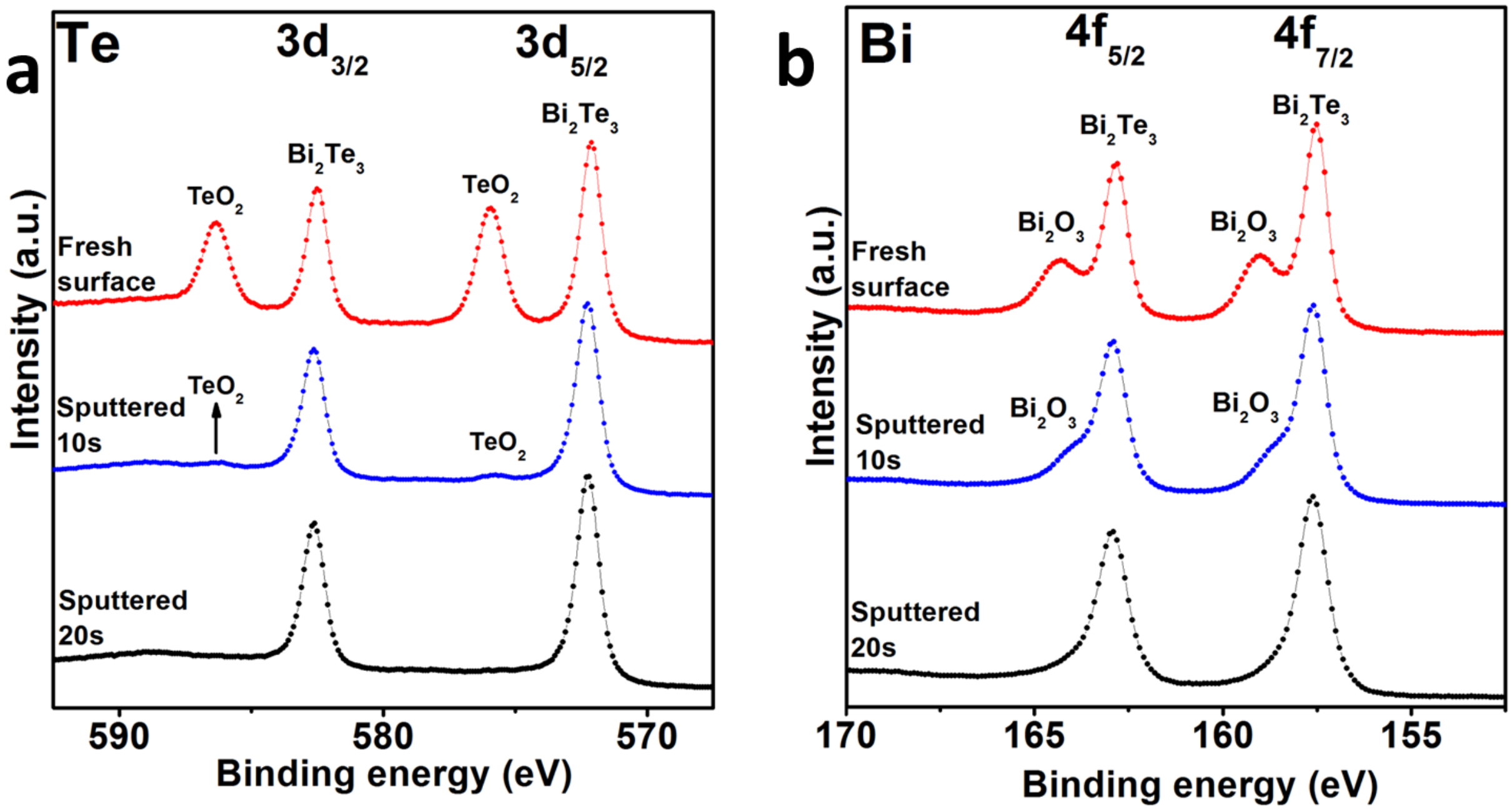

**Supplementary Fig. 8|** Raw data of Te *3d* (a) and Bi *4f* (b) XPS spectra of the 48 nm $Bi_2Te_3$ thin film for its fresh surface, and those after sputtered for 10s and 20s respectively.

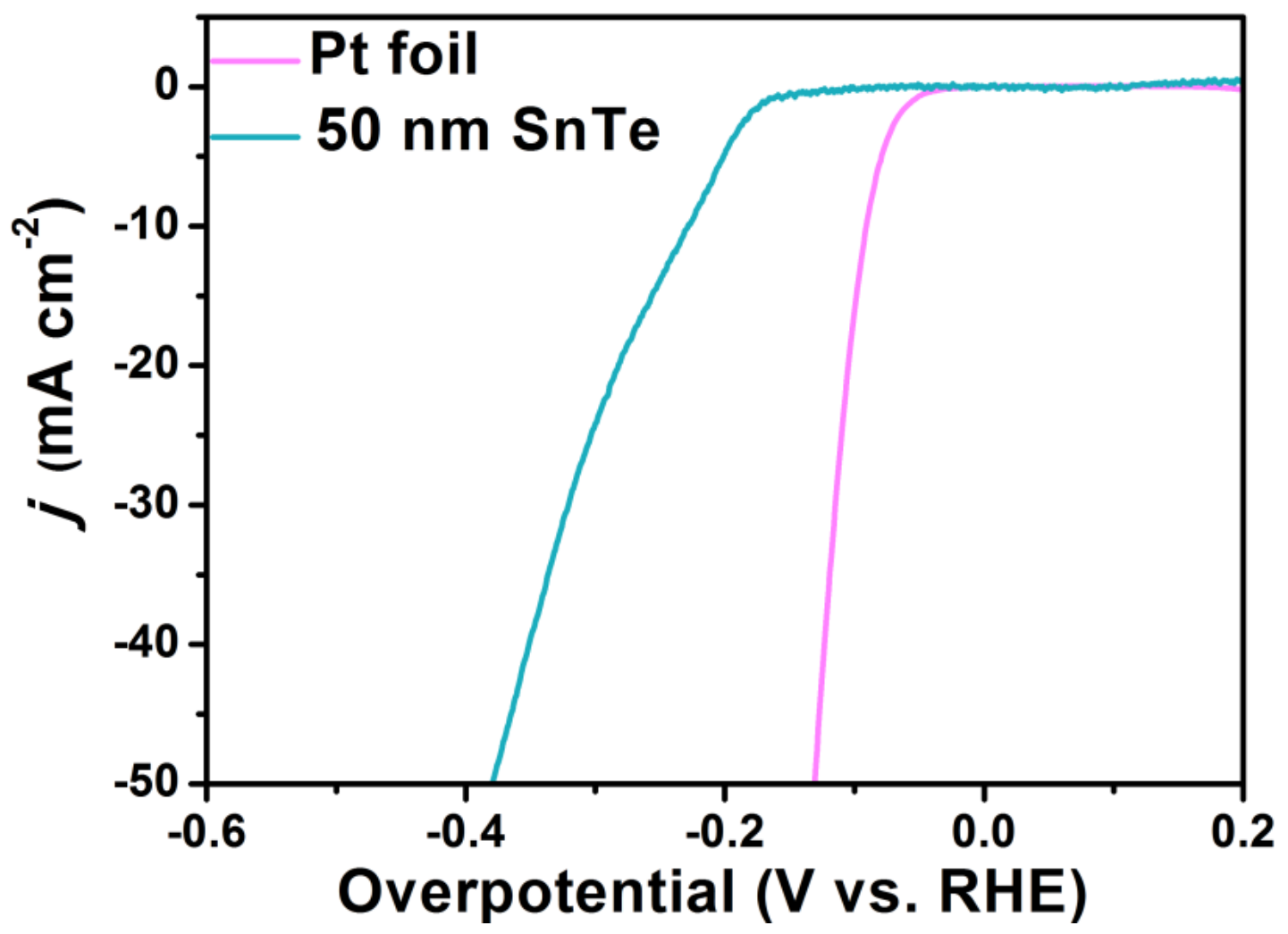

**Supplementary Fig. 9** | Polarization curves (*iR*-corrected) of a 50nm-SnTe thin film and a commercial Pt foil.

**Supplementary Table 1 |** Comparison of the $j_0$ of $Bi_2Se_3$, $Sb_2Te_3$, $Bi_{0.5}Sb_{1.5}Te_3$ nanosheets, exfoliated $Bi_2Te_3$ nanosheets, some $MX_2$-based nanosheet catalysts and this work in Table 1. The electrolytes are all 0.5M $H_2SO_4$.

| Catalyst | $j_0$ (μA cm$^{-2}$) | log $j_0$ (A cm$^{-2}$) |
|---|---|---|
| $Bi_2Te_3$ film (48 nm) (this work) | 1.74 | -5.759 |
| $Bi_2Se_3$ nanosheets [31] | N/A | N/A |
| $MoSe_2$ nanosheets [31] | N/A | N/A |
| $Bi_2Se_3$+$MoSe_2$ nanosheets mixture [31] | N/A | N/A |
| $MoSe_2$/$Bi_2Se_3$ nanostructured hybrids [31] | N/A | N/A |
| $Sb_2Te_3$ nanosheets [32] | N/A | N/A |
| $Bi_{0.5}Sb_{1.5}Te_3$ nanosheets [32] | N/A | N/A |
| $Bi_{0.5}Sb_{1.5}Te_3$ nanosheets on carbon paper (temperature gradient=90 °C) [32] | N/A | N/A |
| Exfoliated $Bi_2Te_3$ nanosheets [26] | N/A | N/A |
| Double-gyroid $MoS_2$ (Mo electrodeposition time: 1min) [27] | 0.69 | -6.161 |
| Vertically aligned homostructured $MoS_2$ nanosheets [30] | N/A | N/A |
| Fewer layer 1T phase $MoS_2$ [30] | N/A | N/A |
| 2H phase $MoS_2$ nanosheets [28] | 3.5 | -5.456 |
| 1T phase $MoS_2$ nanosheets [28] | 3.2 | -5.495 |
| Monolayer $MoS_2$ after desulfurization [29] | 7.9 | -5.102 |

**Supplementary Table 2 |** The EIS fitting results of $Bi_2Te_3$ films, ZnSe buffer and n+GaAs substrate from Nyquist plot.

| | $R_s$ (Ω cm$^2$) | $C_{dl}$(F cm$^{-2}$) | $R_{ct}$ (Ω cm$^2$) |
|---|---|---|---|
| n+GaAs | 3.25 | $2.89\times10^{-7}$ | 1469 |
| ZnSe buffer film | 3.12 | $5.10\times10^{-7}$ | 1065 |
| $Bi_2Te_3$ (5 nm) | 2.118 | $4.649\times10^{-5}$ | 17.33 |
| $Bi_2Te_3$ (17 nm) | 2.012 | $4.892\times10^{-5}$ | 16.48 |
| $Bi_2Te_3$ (34 nm) | 2.035 | $5.313\times10^{-5}$ | 10.22 |
| $Bi_2Te_3$ (48 nm) | 2.111 | $1.466\times10^{-4}$ | 3.744 |
| $Bi_2Te_3$ (61 nm) | 2.049 | $5.927\times10^{-5}$ | 8.266 |